\PassOptionsToPackage{english}{babel}

\documentclass[groupaddress,altaffillsymbol,aip,rsi,twocolumn,reprint,floatfix]{revtex4-1}

\usepackage[USenglish]{babel} 

\usepackage[ansinew]{inputenc}

\usepackage{lmodern} 
\usepackage{float}
\usepackage{graphicx} 

\usepackage{amsmath}

\newcommand{\uv}[1]{\mathbf{\hat{#1}}}

\usepackage{amssymb}
\usepackage{amsthm}
\usepackage{amsfonts}
\usepackage{color}
\usepackage{natbib} 
\usepackage{csquotes}
\usepackage{bm}

\begin{document}

\title{Characterizing atomic magnetic gradiometers for fetal magnetocardiography}
\author{I. A. Sulai} 
\email{ibrahim.sulai@bucknell.edu} 
\affiliation{Department of Physics \& Astronomy, Bucknell University, Lewisburg, Pennsylvania 17837 USA}

\author{Z. J. DeLand} 
\altaffiliation{current address: EPIC Systems Corp. Verona, WI}

\author{M. D. Bulatowicz}
\author{C. P. Wahl}
\altaffiliation{Current Address: AS\&T. UC Berkeley}

\affiliation{Department of Physics, University of Wisconsin-Madison, Madison, Wisconsin 53706 USA} 
\author{R. T. Wakai}
\affiliation{Department of Medical Physics, University of Wisconsin-Madison, Madison Wisconsin 53706 USA}
\author{T. G. Walker}
\affiliation{Department of Physics, University of Wisconsin-Madison, Madison, Wisconsin 53706 USA} 

\date{\today} 

\begin{abstract}

Atomic magnetometers (AMs) offer many advantages over superconducting quantum interference devices (SQUIDs) due to, among other things, having comparable sensitivity while not requiring cryogenics. One of the major limitations of AMs is the challenge of configuring them as gradiometers. We report the development of a spin-exchange relaxation free (SERF) vector atomic magnetic gradiometer with sensitivity of $3\, \rm{fT} \,{cm}^{-1}{Hz}^{-1/2}$ and common mode rejection ratio (CMRR) $> 150$ in the band from DC to 100 Hz. We introduce a background suppression figure of merit for characterizing the performance of gradiometers. It allows for optimally setting the measurement baseline, and for quickly assessing the advantage, if any, of performing a measurement in gradiometric mode. As an application, we consider the problem of fetal magnetocardiography (fMCG) detection in the presence of a large background maternal MCG signal. 

\end{abstract}

\maketitle

\section{Introduction}
 \label{intro}
Magnetic gradiometers can often achieve higher fidelity detection of localized magnetic field distributions than magnetometers.  This is true in environments where the primary noise sources are far from the detectors such as in magnetocardiography (MCG), magnetoencephalography (MEG) \cite{Colombo2016, Sander2012}, geomagnetism \cite{Keenan2011}, ordnance detection \cite{ordnance2} and in low field NMR \cite{Jiang2018}, with the benefit arising with suppression of common mode noise. Sensitivity, bandwidth and dynamic range determine the optimal choice of sensor for each application.
 
 Consider a situation where a \textit{uniform} magnetic field $B_c$ component (or magnitude for scalar sensors) is measured by two closely separated sensors. The common mode rejection ratio, CMRR is
\begin{equation}
\label{eq:CMRRdef}
\xi = \frac{M_1 B_{1c} + M_2 B_{2c}}{2(M_1 B_{1c} - M_2 B_{2c})} =\frac{M_1 + M_2}{2(M_1  - M_2 )} 
\end{equation}
where $B_{ic}$ is the field measured by sensor $i$, and $M_i$ is the transfer function for each sensor. For identical sensors in this situation $(M_1= M_2,B_{1c}=B_{2c})$, we expect a CMRR of infinity resulting in vanishing sensitivity to the difference signal. In practice however, differences in sensors lead to a finite CMRR $\xi$.

Presently, the highest performing magnetic gradiometers involve inductively coupling the magnetic field of interest to low Tc superconducting quantum interference devices, SQUIDs via counter-wound pickup coils \cite{Quartero2002,Strasburger2008}. The SQUID gradiometer CMRR  is limited by the degree to which  the areas and alignments of these (often hand-wound) coils are matched. Typical values of SQUID gradiometer balance is about one part in 100, setting the upper bound of the common-mode suppression at that level. \cite{Finland1993,Clarke2004}. 

Unlike SQUIDs, atomic magnetometers (AMs)  require no cryogens. They are relatively inexpensive to operate and have significantly lower setup and operating costs due to their working temperatures and smaller shielding volume requirements. However, configuring AMs as gradiometers poses a number of challenges and is an active area of work \cite{Affolderbach,Dang,Kominis,Kim,Kamada,Sheng,Sheng2017,Johnson,Shoujun}. 

Most AM gradiometer implementations, this work included, involve subtracting magnetic field measurements from adjacent sensors in post processing. In this paper, we consider the challenge of performing such a subtraction and recommend procedures for calibrating and characterizing AM gradiometers. In section \ref{sec:FOM}, we introduce a figure of merit for evaluating gradiometer performance.  It is a function of the CMRR, the gradiometer baseline, as well as the geometric scaling of the signal and the dominant background fields. This metric will be useful for comparing different implementations of gradiometers when applying the sensors to a particular measurement problem.  In section \ref{sec:SERFgradio}, we report the development of Spin Exchange Relaxation Free (SERF) gradiometers with performance metrics comparable to SQUIDs. Finally, in section \ref{sec:FMCG}, we consider the problem of  detecting fetal magneto-cardiography (fMCG) in the presence of the large maternal MCG signal. Our goal is to efficiently isolate the fMCG signal from the maternal background. Some relevant questions to address are how best to configure the sensors, what CMRR is sufficient and how to compare different gradiometer implementations. We shall discuss the gradiometer we have developed in comparisons with others in the literature.

\section{Gradiometry Figure of Merit}
\label{sec:FOM}
Often, measurements from an array of magnetic field sensors are input into an analysis pipeline for further processing: e.g for source localization, Independent Component Analysis etc. The question we consider is if and when does using the output of gradiometers provide an advantage over using magnetometers? 

Consider a magnetic field signal described by the power law $B(r) \propto (1/r)^p$ in the presence of noise with rms amplitude $\delta B = (\delta B_u^2 + \delta B_c^2)^{1/2}$, where $\delta B_u$ and $\delta B_c$ are the rms magnitudes of uncorrelated and correlated (uniform) noise respectively. A gradiometric measurement of such a field is only beneficial if the signal-to-noise ratio in gradiometric mode is superior to that obtained from two separate magnetometers. We therefore define a figure of merit $\mathcal{F}$ as the ratio:
\begin{align}
\label{eq:FOM1}
\mathcal{F} = \frac{SNR_G}{SNR_M}	&= \frac{g^p-1}{g^p+1}\frac{\sqrt{\left(\delta B_u\right)^2+ \left(\delta B_c\right)^2}}{\sqrt{\left(\delta B_u\right)^2+\left(\frac{\delta B_c}{\xi}\right)^2}},
\end{align}
where $g \left(= 1 + L/r_s\right)$ is a dimensionless parameter which includes the \enquote{baseline} distance $L$ between the two detectors and source distance $r_s$. (Full derivation is given in Appendix \ref{App:FOM}). The contribution of the correlated noise is suppressed by a factor of the CMRR.

Understanding the spatial characteristics of the dominant background is important when considering gradiometers for a measurement. We identify a number of limiting cases of Equation \ref{eq:FOM1}.  

\begin{itemize}

\item{Case (0):} ${\delta B_u \gg \delta B_c}$, that is in situations for which the uncorrelated noise dominates, $\mathcal{F}<1$, and there is no advantage to using gradiometers. Doing so will only reduce the SNR.

\item{Case (1)}: $\delta B_u \ll \delta B_c$. Here we then have that
\begin{align}
\label{eq:FOM2}
\mathcal{F} \approx \frac{g^p-1}{g^p+1}\frac{1}{\sqrt{\left(\frac{\delta B_u}{\delta B_c}\right)^2 + \xi^{-2}}}. 
\end{align}
In this regime, a combination of $\xi$ and $\delta B_u \ll \delta B_c$ governs the figure of merit. $\mathcal{F}$ in this case will be a sensitive function of the baseline. Investing in much higher CMRR might not yield significant change. Other strategies might yield higher payoff. 

\item{Case (2)}: $\delta B_u \ll \delta B_c$ and $\left(\delta B_u/\delta B_c\right)^2 \ll \xi^{-2}$,

This describes a scenario where the ambient magnetic background is highly correlated.

\begin{align}
\label{eq:FOM3}
\mathcal{F} = \frac{g^p-1}{g^p+1}\xi
\end{align}
An example of this would be operating the gradiometer in the earth's field, away from strong local sources. This is the ideal situation to operate a gradiometer. The figure of merit in this case is primarily limited by the CMRR. Design decisions which improve the CMRR will yield a high payoff.
\end{itemize}

We note that the geometric factor $(g^p-1)/(g^p+1)$ is a monotonically increasing function of the baseline. Therefore in case (2), larger baselines are desirable. However, in general, the optimal baseline will depend on the length over which the background remains uniform.

\section{Development of SERF Gradiometer}
\label{sec:SERFgradio}
We have developed a diffusion-mode, two-beam SERF magnetometer array configured as a gradiometer for use in fetal magnetocardiography, fMCG \cite{Wyllie,Sulai2013}. SERF conditions require working in a near-zero field environment \cite{Allred}. As a result, we work in a (two-layer mu-metal, 1-layer Aluminum)  magnetically shielded room (MSR). The MSR has a residual DC field on the order of 50 nT which we compensate for using active cancellation via a large ($D \approx$ 3 m) set of 3-axis \enquote{common} coils wrapped on the perimeter of the MSR and sets of small ($D \approx$ 4 cm) \enquote{local} coils wound around each individual magnetometer.

One potential source of magnetic noise is the noise on the currents creating the compensation fields. A 50 nT DC field for example requires a 150 dB$\sqrt{\text{Hz}}$ signal to noise ratio to reach 1 fT/$\sqrt{\text{Hz}}$. We compensate the majority of this DC field with large coils wound near the walls of the room, such that noise associated with these shimming currents is common-mode to a large degree. In practice, we null the field of one of the channels using the large set of coils, and then minimize the residual fields in the other channel ($\leq$ 5 nT) with its set of local coils. This reduces the current noise requirement by 20 dB. In our experiments, the common-mode noise produced by these large sets of coils was measured to be approximately 5 $\text{fT}/\sqrt{\text{Hz}}$ at the location of the magnetometers.

\begin{figure*}[!t]
		\includegraphics[width = 0.9\linewidth]{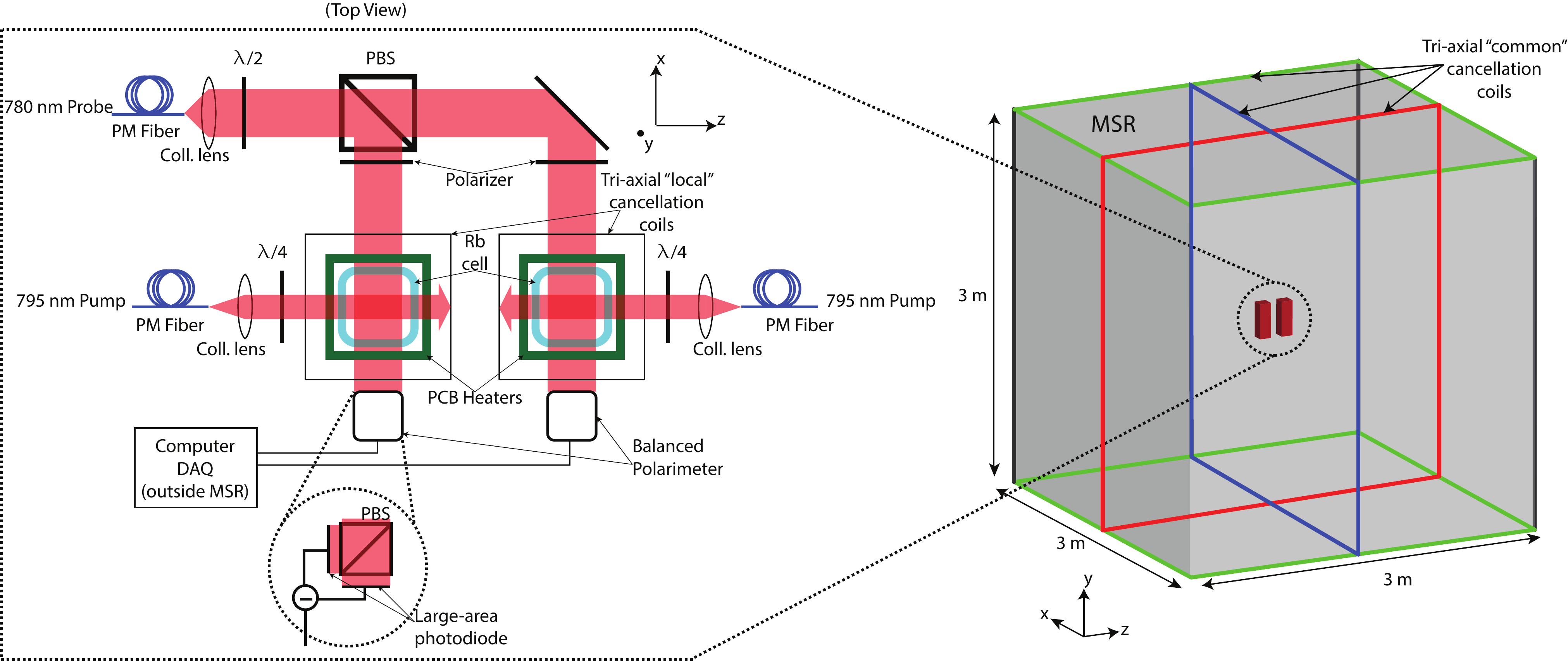}
\caption{ Two SERF magnetometers comprising the gradiometer are placed at the center of a magnetically shielded room. Large (common)coils wrapped around the perimeter of the innermost shield are used to null the total field at one of the magnetometers, as well as to apply calibration signals. Local coils compensate residual gradients. The  magnetometers are a distance $L$ = 4 cm apart. A circularly polarized pump beam polarizes a rubidium vapor along the $\hat{z}$ axis. Magnetic fields along $\hat{y}$ (or $\hat{x}$ and $\hat{y}$ in Z-mode)  cause rotation of the atomic electron spin polarization which is detected by Faraday rotation of an off-resonance linearly polarized probe beam using a balanced polarimeter. }
\label{fig:Apparatus}
\end{figure*}

Two channels (shown in Figure \ref{fig:Apparatus}) are spaced a distance of $L = 4\,\rm{cm}$ apart. Each channel comprises a $1\, \rm{cm}^3$ Pyrex vapor cell with Rubidium and approximately $200 $ Torr of Nitrogen and Neon buffer gas. We heat the cells up to a temperature of  $150 ^\circ$ C, elevating the vapor pressure, such that the equilibrium density is $n = 10^{14}\, \rm{cm}^{-3}$. The basic idea behind the operation of the SERF magnetometer is that an electron spin polarization $\mathbf{P}$ established by optical pumping interacts with the ambient field, and relaxes at a rate governed primarily by background atom collisions. Rubidium atoms are polarized along $\mathbf{z}$ with circular polarized light resonant with the Rb D1 line at 795 nm. We subsequently detect the projection of $\mathbf{P}$ along the probe direction, $P_x$, via off-resonance Faraday rotation using light near the Rb D2 line at 780 nm.

The spin dynamics of the polarization are described by the Bloch equations and are discussed more fully in Appendix \ref{sec:AppSERFMag}. 

\subsection{Operating Modes}
\begin{figure}[h!]
	\includegraphics[width = 0.75\linewidth]{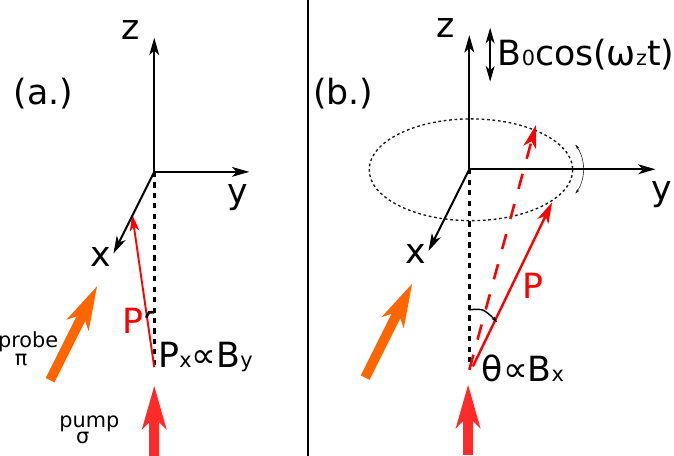}
\caption{Optical pumping along $\uv{z}$ establishes an electron spin polarization. Off resonance Faraday rotation of the probe light senses the projection of the polarization along the $\uv{x}$ axis.  Panel (a.) shows the DC mode for which $P_x(t) \propto B_y(t)$. In panel (b.), we show Z-mode, where a modulating field is applied along z, and  $P_x(t) \propto B_z \, \rm{cos}(\omega_z t) $}.
\label{fig:modScheme}
\end{figure}

We operate the instrument in either a so called `DC mode' or a parametrically modulated `Z-mode.' In the DC operating mode, the component of the magnetic field perpendicular to the plane of the pump and probe lasers ($B_y$) tilts the atomic polarization onto the detection axis \cite{Allred,WyllieThesis}. This maps $B_y(t)$ on to the Faraday rotation signal. See panel (a.) of Figure \ref{fig:modScheme}. 

In Z-mode, we apply a high frequency modulating magnetic field aligned along the pumping direction ($B_0\cos{\omega_z t}$, with $\omega_z \sim 2 \pi \times 1 \,\rm{kHz}$)  \cite{Li}. A field along $\uv{x}$ torques the polarization $\mathbf{P}$ into the $y-z$ plane at an angle proportional to $B_x$. The polarization is  modulated at $\omega_z$ in the $x-y$ plane as shown in panel (b.) of Figure \ref{fig:modScheme}. By demodulating the detected $P_x$ signal at $\omega_z$, we gain sensitivity to $B_x$  at the fundamental frequency (and higher odd harmonics). In addition, we retain sensitivity to $B_y$ at DC (and even harmonics of the modulation) \cite{Li}. This lock-in technique has the advantage of moving the signal away from DC, to a higher frequency which is free of low frequency laser intensity and polarization noise. Therefore in Z-mode, we can in principle simultaneously detect two components of the magnetic field gradient tensor: $\partial{B_y}/\partial{z}$ and $\partial{B_x}/\partial{z}$ per the coordinate system shown in Figure \ref{fig:Apparatus}. For this work, we report measurements where only one component of the tensor was considered at a time.

The DC and Z-mode sensors can be run in open-loop, described above, or in closed-loop, where we use a PID circuit to drive the measured field to zero in the sensitive direction at each sensor position. The compensating signal applied by the servo is then equal to the time varying magnetic field.  One of the advantages of running in closed-loop mode is an increased dynamic range. This arises from keeping the average field at the sensor location near zero -- thus avoiding saturating the detectors and/or amplifiers. We observe that the closed loop dynamic range is $\sim 100 \,\rm{nT}$ compared to $\sim 10 \,\rm{nT}$ in the open loop mode. Second, maintaining the field near zero keeps the polarimeter signal balanced. Residual laser intensity fluctuations on the probe can then be suppressed to a higher degree \cite{Hobbs}. Thirdly, the feedback mode flattens the response of the device through design of the feedback circuit. Operating in closed-loop, we are able to broaden the 3 dB bandwidth of the magnetometers from $\sim 50$ Hz to $\sim 120 $ Hz.

\subsection{Calibration / Characterization}
Due to differences in vapor cell gas composition, temperature, pump and probe laser characteristics etc, the two magnetometers may have different amplitude and phase responses. Their signals cannot be simply subtracted. The full complex response of the two sensors must be properly accounted for. Failure to do so will result in poor common mode rejection due to dephasing errors.  Because of this we calibrate the magnetometers against some common calibration signal which we design to ensure adequate coverage of the whole band of interest, as  inadequate SNR in the calibration step can limit the achievable CMRR.

\begin{figure}[h!]
	\includegraphics[width = 0.9\linewidth]{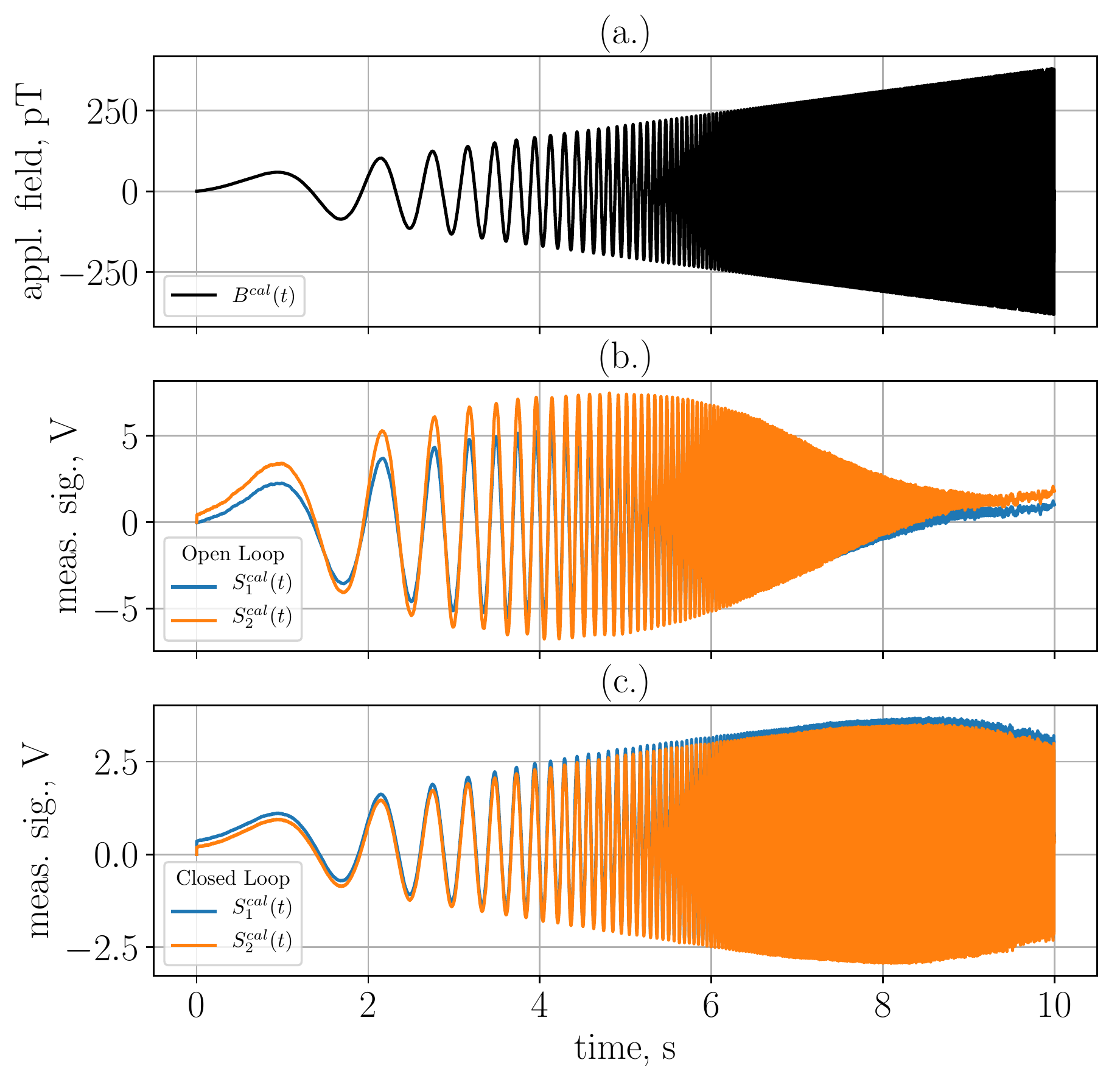}
\caption{Two magnetometer channels comprising a gradiometer  were calibrated simultaneously by applying a time dependent field of the form of equation \ref{eq:cal2} to a common coil shown in (a.). Panel (b.) shows Z-mode open loop response to the applied fields. (c.) shows the closed loop response where the increased bandwidth is clearly evident}.
\label{fig:gCalib}
\end{figure}

We apply a signal $B^{cal}(t)$ to the large coil and obtain calibration responses from the two channels $S^{cal}_1(t)$ and $S^{cal}_2(t)$.  We choose calibration fields to be of the form 
\begin{equation}
\label{eq:cal2}
B^{cal}(t) = \left( \alpha_0 + \alpha_1 t \right)\, \sin (\omega(t)\, t ),
\end{equation}
 where $\omega(t) = \alpha_2 \exp{\alpha_3 t}$, with $\alpha_i$ positive constants which govern the `speed' of the frequency chirp. This function has an exponentially increasing frequency and a linearly increasing amplitude, which ensures that the SNR for the calibration signal and response does not limit $\xi$ in the DC to 200 Hz frequency range. The applied calibration signal along with the magnetometer responses are shown in Figure \ref{fig:gCalib}.
\begin{figure}[h!]
		\includegraphics[width = \linewidth]{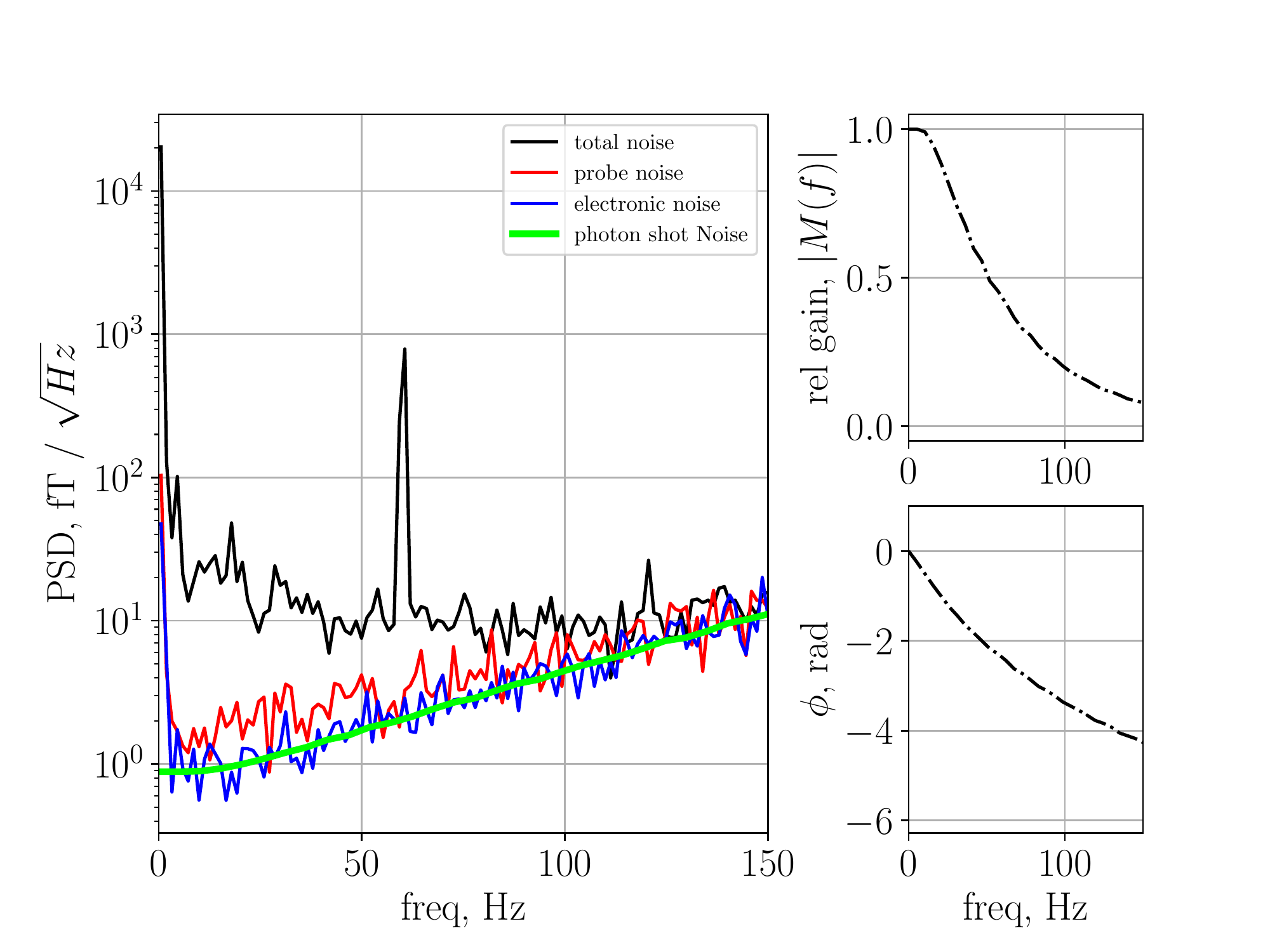}
\caption{Calibrated response of a single channel of the open-loop magnetometers used in constructing the gradiometer. The total noise is dominated by the magnetic noise in the MSR.  The amplitude and phase response is shown on the right.}
\label{fig:Response}
\end{figure}

By taking the Fourier transforms of $B^{cal}(t)$ and $S^{cal}(t)$, we obtain the transfer function of each sensor: $ \tilde{M}_i(f) =\tilde{S}^{cal}_i(f)/\tilde{B}^{cal}(f)$. We interpolate the measured transfer function -- obtaining a functional representation of the complex gain which we use to transform subsequent measurements. On acquiring the time series $S_1(t)$ and $S_2(t)$, we compute a gradiometric signal (in $\rm{fT/cm}$) by calculating
\begin{equation}
\label{eq:cal1}
\begin{split}
G(t) &= (B_1(t) - B_2(t))/L \\
     &= \mathcal{F}^{-1}\left\{ \tilde{B}^{cal}(f) \left(  \frac{\tilde{S}_1(f)}{\tilde{S}^{cal}_1(f)} -  \frac{\tilde{S}_2(f)}{\tilde{S}^{cal}_2(f)}  \right)  \right\}\times\frac{1}{L}\\
     &= \mathcal{F}^{-1} \left\{   \frac{\tilde{S}_1(f)}{\tilde{M}_1(f)} -  \frac{\tilde{S}_2(f)}{\tilde{M}_2(f)}    \right\} \times\frac{1}{L}
\end{split}
\end{equation}
where $\mathcal{F}^{-1}$ represents an inverse Fourier transform and $L$ is the gradiometer baseline. We have assumed here that the transfer functions $M_i(f)$ are stable on the calibration and measurement time scales. Furthermore, we use the results of the fit to $M_i(f)$ to design control servos \cite{DeLandThesis} for closed loop operation.

In Figure \ref{fig:Response}, we plot the complex response of a single magnetometer. The magnetometers' (DC normalized) relative gain $M(f) = |\tilde{M}(f)|/|\tilde{M}(0)|$ and phase dependence are shown on the two right panels.  We also plot the noise spectral density scaled into magnetic field units. The black trace represents the total effective magnetic ambient noise in the room. The probe noise is the noise associated with the intensity and laser polarization fluctuations of the probe laser and has a lower bound given by the photon shot noise.
\begin{figure}[b!]
\centering
\includegraphics[width = \linewidth]{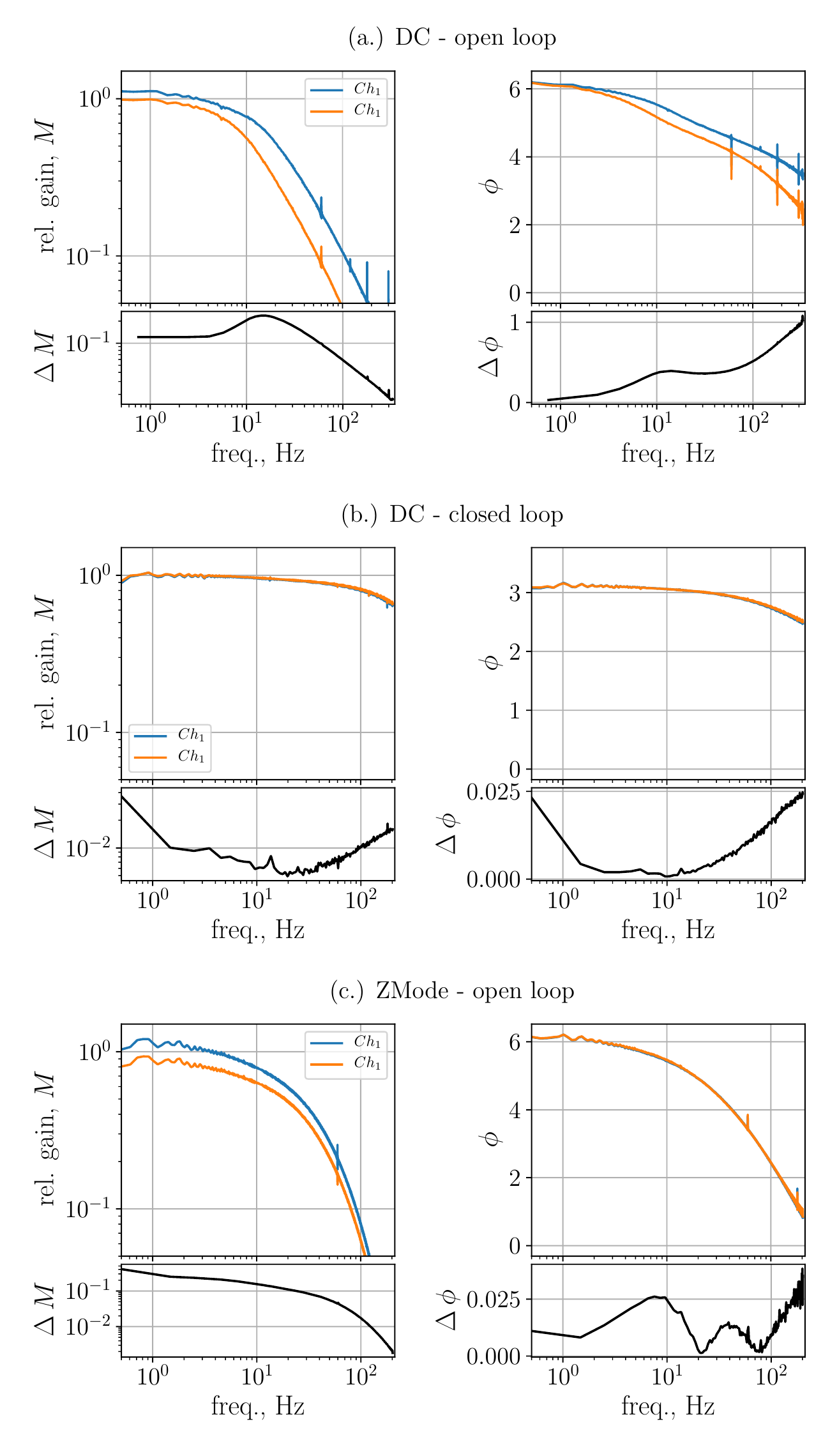}
\caption{Complex response function of the different operating modes. The lower plots in each of panels (a.) to (c.) show the gain and phase difference $\Delta M$ and $\Delta \phi$ of two magnetometer channels. These differences are minimized for an optimized configuration. }
\label{fig:complexResp}
\end{figure}
We measure the probe noise by observing the signal fluctuations with the pump light blocked. Similarly, we measure the electronic noise -- the noise due to digitization and EM pickup on the DAQ input lines  by blocking both the pump and probe light. In the situation shown in the figure, the magnetic noise dominates all other noise sources. This establishes a margin within which a gradiometer can suppress the uniform component of the magnetic noise. This margin can be amplified by increasing the overall gain of the magnetometer -- e.g. by using larger volume cells, or multi-pass cells \cite{Sheng,Li}. If the individual magnetometers are limited by technical noise, i.e. probe or electronic noise, then there is no advantage to operating in gradiometric mode. This corresponds to case (0) described in section \ref{sec:FOM}.

In Figure \ref{fig:complexResp}, we show the relative gains of pairs of magnetometers comprising a gradiometer. Panels a - c show DC - open loop implementation, DC - closed loop, and Z-Mode open loop implementations respectively. We plot the difference in response between the two sensor $\Delta M$ and $\Delta \phi$ as the lower plots of each panel. The phase and amplitude response of the two sensor is particularly striking in the open loop operation mode. If not properly accounted for, dephasing errors proportional to $1 - \cos{\delta \phi}$, where $\delta \phi$ is the error on $\Delta \phi$, would show up in the difference channel and subsequently in the CMRR. The idea here is that the two channels are vectors in the complex plane, and the dephasing error is the magnitude of the difference vector not accounted for by the calibration. 

Panel (b.) shows the measured response of the two gradiometers in DC-mode, closed-loop. Notice the expanded bandwidth, as well as the diminished phase difference. This portends well for achieving high CMRR. Indeed we find that to be the case. The responses are very similar because in closed loop, the shape of the response is dominated by the servo electronics. 

Panel (c.) shows the Z-mode open-loop response. One prominent feature of this mode is that the phase response of the two sensors is dominated by the low pass filter associated with the demodulation. Because of the steep phase response we could not, using the same PID loop, apply stable negative feedback over the 0 - 200 Hz band to a Z-mode closed-loop sensor as we did in the the DC-mode closed loop case. We therefore do include Z-mode closed-loop case in our comparisons. Implementing this is on the agenda for future work. We believe however, that the major conclusions of this work stand without it. 

\begin{figure}[!h]
\includegraphics[width = \linewidth]{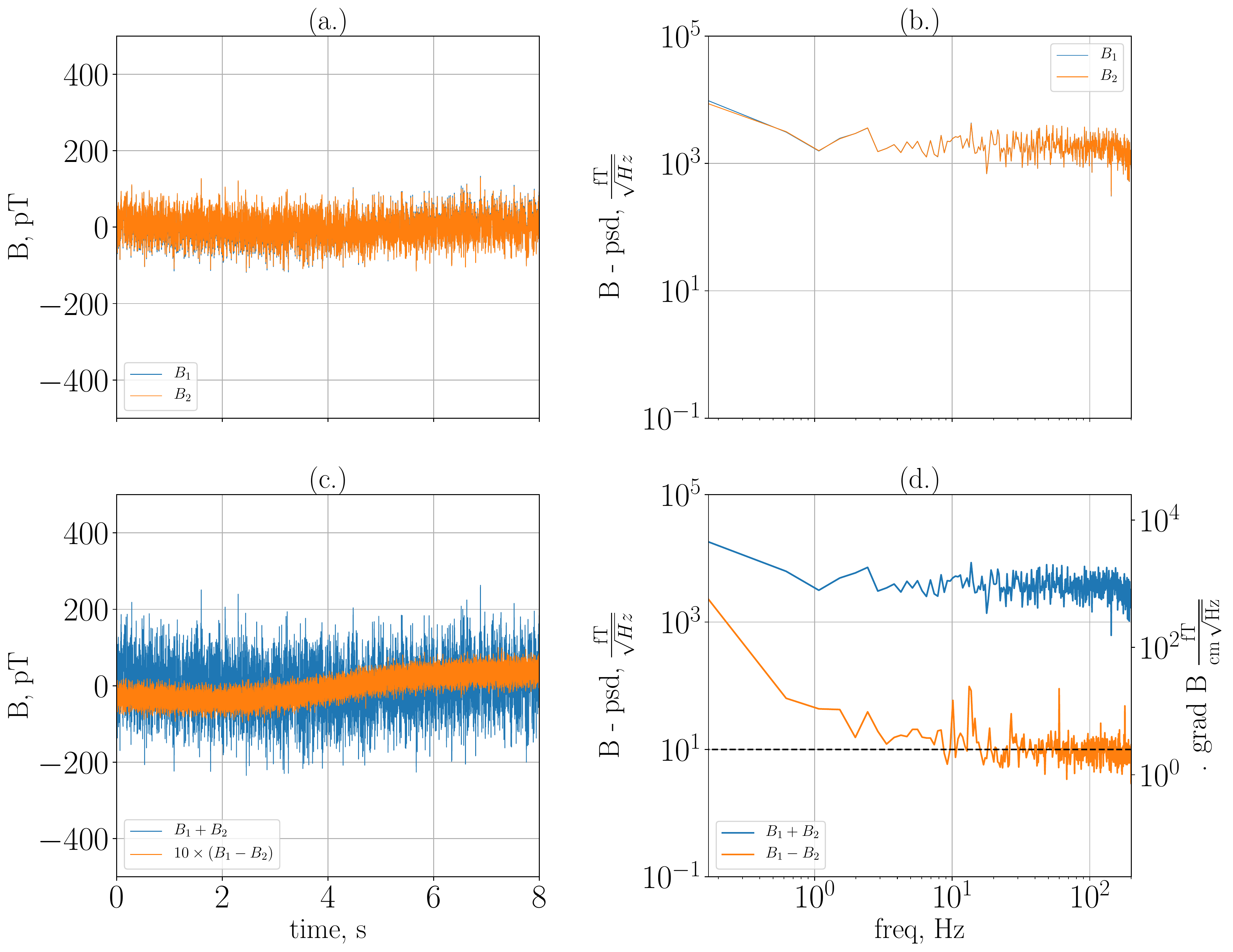}
\caption{Determination of CMRR, $\xi$. White noise is added to the magnetically shielded room via the large coils mounted on the inside surface of the MSR. Panels (a.) and (c.) show  the time domain measurements of the individual sensors, as well as their sum and difference. Panels (b.) and (d.) show the power spectral density of the individual as well as of the combined channels. We estimate $\xi$ from the measurements in panel (d). The dotted line is at 2.5 fT/$\sqrt{Hz}\mathrm{cm}$  as a guide for the eye. }
\label{fig:CMRRcalc}
\end{figure}

\begin{figure}[h!]
	\centering
		\includegraphics[width=\linewidth]{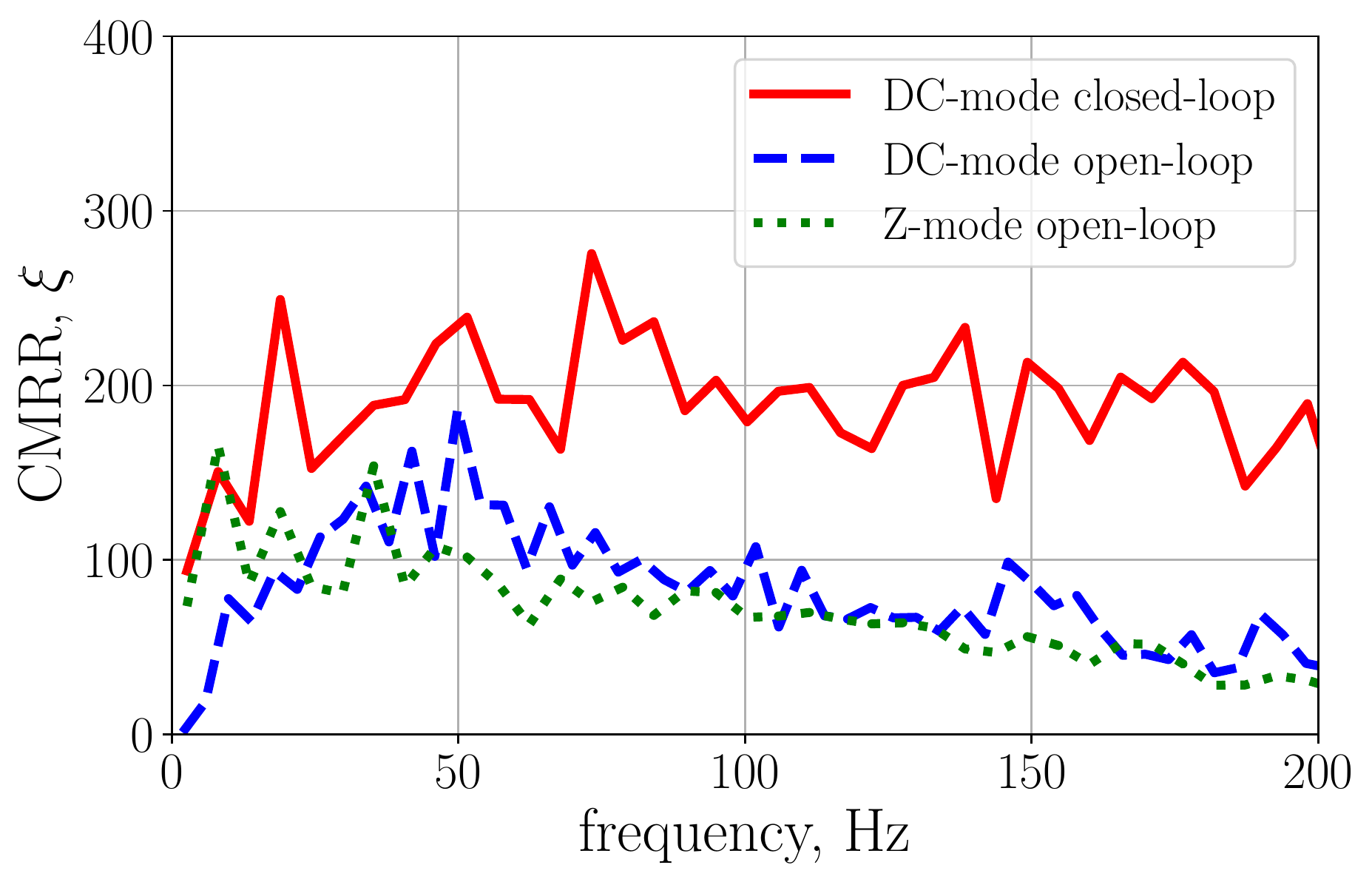}
\caption{ Comparison of the CMRR $(\xi)$ for various operating modes. $\xi$ was determined by using equation \ref{eq:CMRRdef} for measurements made with applied noise as in Figure \ref{fig:CMRRcalc}}
\label{fig:CMRRComparo}
\end{figure}

We determined the common-mode rejection ratio $\xi$ of the gradiometer described above by measuring the suppression factor of digitally synthesized white magnetic noise applied to the large coils.  It is critical in this step that the amplitude of the applied noise be greater than $\xi$ times the noise floor of the sensors. Otherwise $\xi$ will be underestimated.

In Figure \ref{fig:CMRRcalc}, we show CMRR measurement of the DC - closed loop gradiometer. We observe noise in the difference channel of $< 12 fT/\sqrt{Hz}$ over a baseline of 4 cm. This corresponds to a magnetic field gradient noise of $< 3 fT/\mathrm{cm} \sqrt{\mathrm{Hz}}$ In Figure \ref{fig:CMRRComparo}, we show a compilation of the measured CMRR for the three operating modes. We found that the DC closed-loop gradiometer had the highest CMRR, with the two open loop cases being comparable. This is summarized in  Figure \ref{fig:CMRRComparo}. 

In summary, we have developed a closed loop first order gradiometer with CMRR $\xi\sim 150$, and gradient noise  $ < 3\, \rm{fT\,cm^{-1}} /\sqrt{\rm{Hz}}$.

\begin{figure}[h!]
	\centering
		\includegraphics[width = \linewidth]{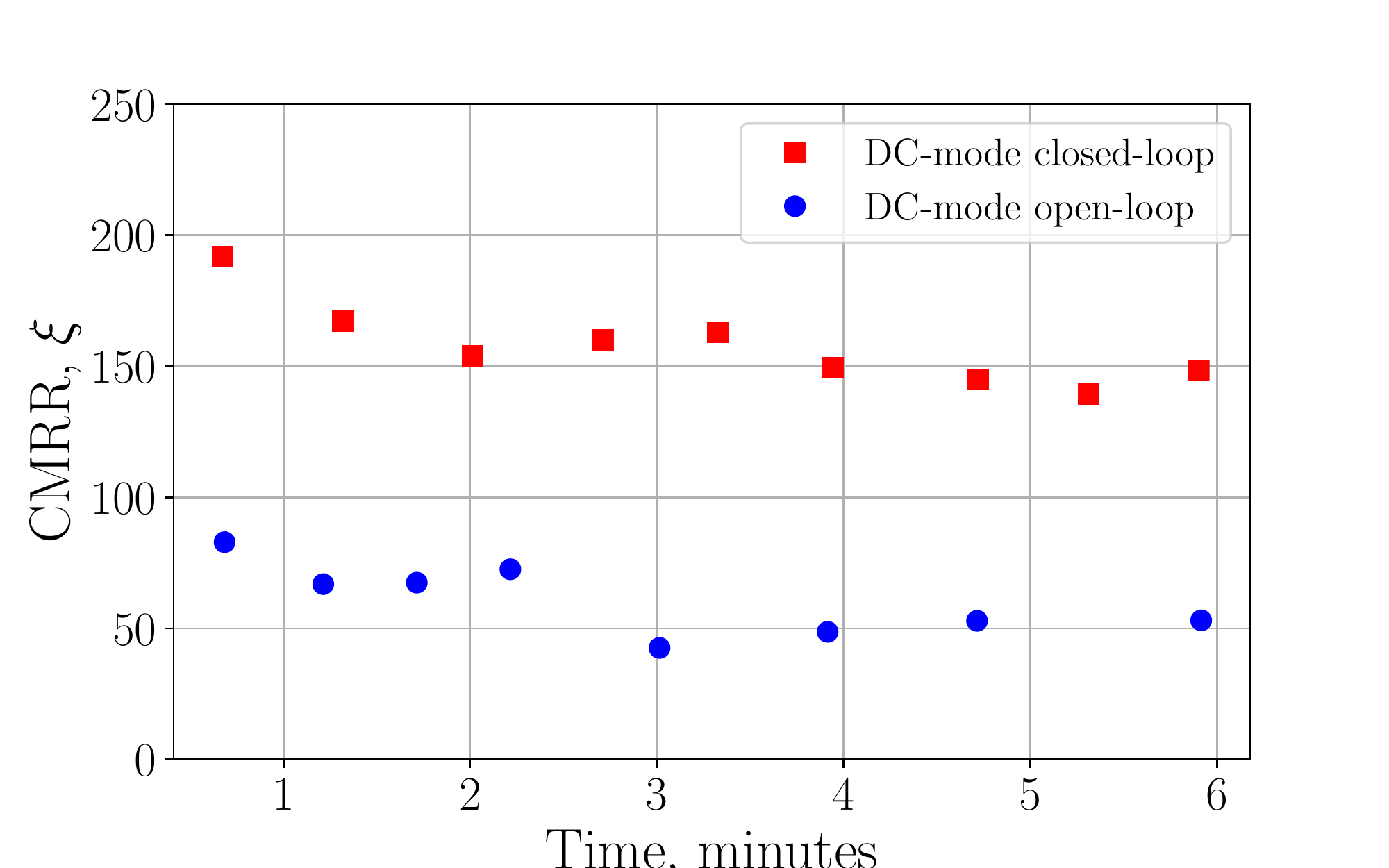}
\caption{The stability of $\xi$ over time was measured in DC-mode with and without active feedback. In the former case, the device was able to maintain an acceptably high level of common-mode noise suppression over a time period comparable to a fMCG measurement. When operating without the feedback, however, the device suffers both in the absolute value of $\xi$ as well as its stability.}
\label{fig:CMRRStability}
\end{figure}

\section{Gradiometers for fMCG}
\label{sec:FMCG}
 As noted by Hornberger et al., \cite{HORNBERGER2008}, ``Fetal magnetocardiography (fMCG), the magnetic analog of fetal ECG, is at this time the most effective means of assessing fetal rhythms.'' It enables the in-utero diagnoses and monitoring of congenital heart abnormalities \cite{Batie}. The primary technical challenge of fMCG arises from the relative weakness of the fMCG signal in comparison to background fields -- especially the maternal MCG. 

Presently, the FDA approved instruments for clinical fMCG applications are SQUID arrays comprising on the order of  $\sim 20$ gradiometers, \cite{FDAdoc,Clarke2004}. This multiplicity of channels offers an advantage of higher order spatial filtering, or equivalently, the implementation of higher order gradiometers. SQUID based fMCG measurements require magnetically shielded rooms of sufficient size to accommodate the large liquid helium dewars needed for cooling the devices. The required size of the shielded rooms, and the necessary cryogenics, make SQUID gradiometers expensive to set up and operate. 

\subsection{MCG signal properties}

The magnetic field from a heart is often modeled as that arising from a current dipole\cite{Cohen1979},  $B(r_s) \propto (1/r)^p$, with  $p \sim 1.8-2.2$ . Consequently, with $g = 1+ L/r_s$, we calculate the the figure of merit for detecting fMCG using equation \ref{eq:FOM2}. Given the baseline $L$, and distance from the source $r_s$, we choose an optimal sensor for fMCG to be that which maximizes the figure of merit given in equation \ref{eq:FOM2}.

\begin{figure}[b!]
	\centering
		\includegraphics[width = \linewidth]{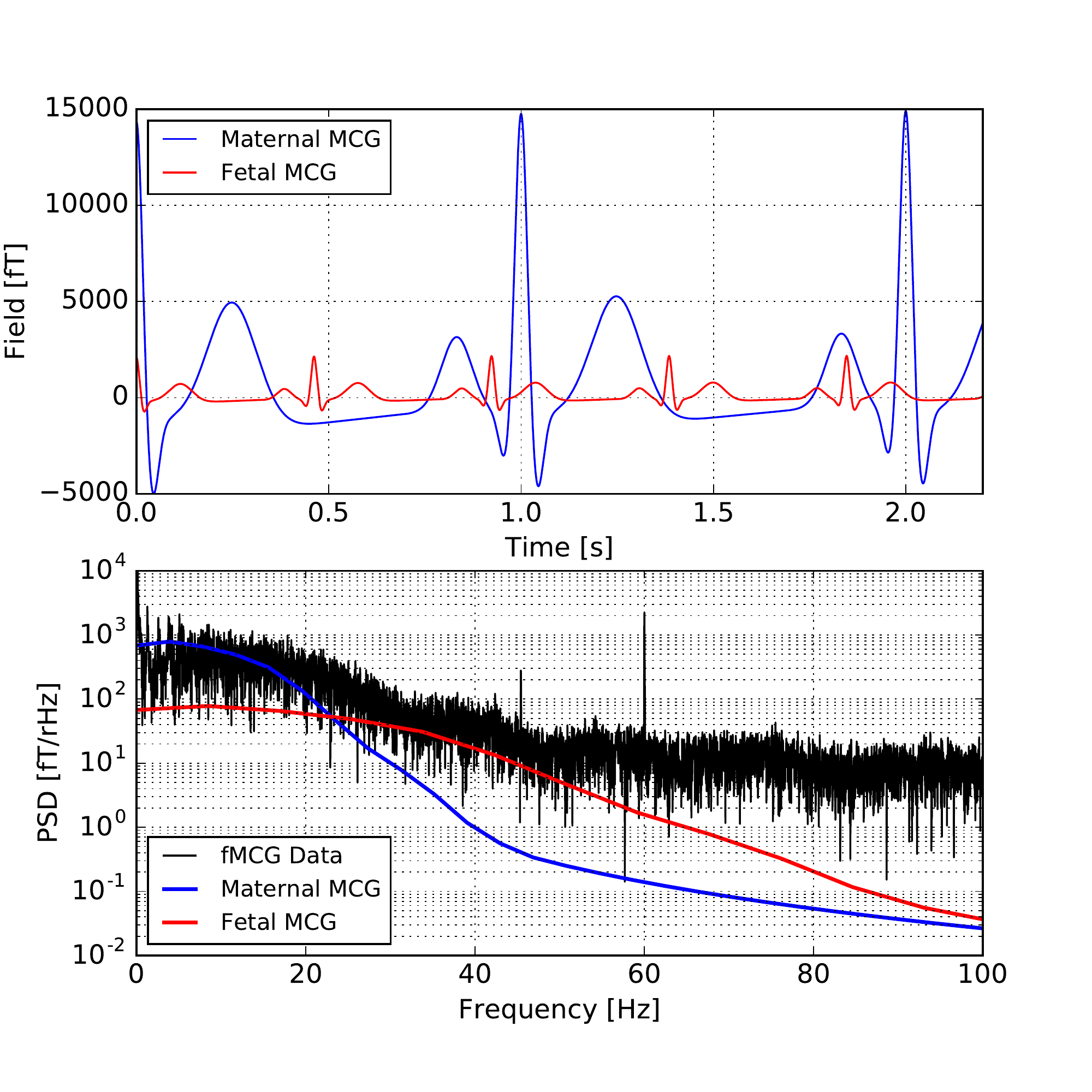}
\caption{Simulated fMCG and MCG wave forms using the ECGSYN program. The power spectral density from the simulation, along as from  actual data collected from a recent fMCG data collection run is also shown. Sensors with a bandwidth of $\sim 100 \, \mathrm{Hz}$ and sensitivity of a few  $\sim 10 \mathrm{fT/\sqrt{Hz}}$ will detect  the majority of the signal.}
\label{fig:ECGSYN}
\end{figure}

In assessing the interference from the maternal background on the fMCG, we simulated MCG wave forms using the ECGSYN \cite{ECGSYN} program -- scaling the field strengths to the magnitudes typically observed. As shown in Figure \ref{fig:ECGSYN}, the maternal signal is roughly ten times that of the fetus \cite{WyllieThesis}. This corresponds to an arrangement where the maternal heart signal originates $\sim 60$ cm from the device (compared to the $\sim 5$ cm distance to the fetal heart). We compute and plot in Figure \ref{fig:ECGSYN} the power spectral density of the two signals, and compare them with the spectral density from a real fMCG measurement obtained using our SERF magnetometers. From this, we posit that for fMCG detection, magnetometers with sensitivity of a few $fT/\sqrt{Hz}$ and bandwidth $\sim 100 Hz$ are necessary. 

Typically also, fMCG measurements acquired over a ${1\,\rm{to}\,3}$ minute window can be averaged in order to determine relevant MCG parameters such as the time intervals between phases of the MCG waveform. We therefore desire sensors whose calibrations are stable over the course of a measurement interval that is averaged. Finally, the CMRR of the gradiometer will serve to suppress ambient magnetic noise that is uniform. The degree to which it helps suppress the maternal MCG will be described below.

The AM gradiometer we have described above meets the stated sensitivity and bandwidth  requirement, along with having $\xi > 100$ over the relevant band. It is also stable over the averaging time. We tested this by performing a calibration after which we measured the CMRR over a period of $\sim$ 5 minutes, as shown in Figure \ref{fig:CMRRStability}. The average value of $\xi$ over the band of 1-100 Hz was recorded at various times after calibration in both \enquote{open-loop} and \enquote{closed-loop} modes. The stability of the CMRR over this interval suggests that the calibration is sufficiently stable for the signals to be averaged. As in previous experiments, we find $\xi$ was superior when operating in closed-loop mode. We did however observe a measurable reduction of $\xi$ with time, and hypothesize that the observed degradation is dominated by residual sensitivity to transverse fields. In our setup, the feedback was only applied along one direction. We suspect that drifts of the field in the other two  directions could change the complex response of the two channels. Implementing a scheme where the fields are sensed and compensated in all three directions is one of our priorities for future studies. Operating in closed loop, we find that we can comfortably operate with $\xi > 100$.

\subsection{On isolating fetal signal from maternal background}
\label{sec:MCGsep}
\begin{figure}[h!]
	\centering
		\includegraphics[width = \linewidth]{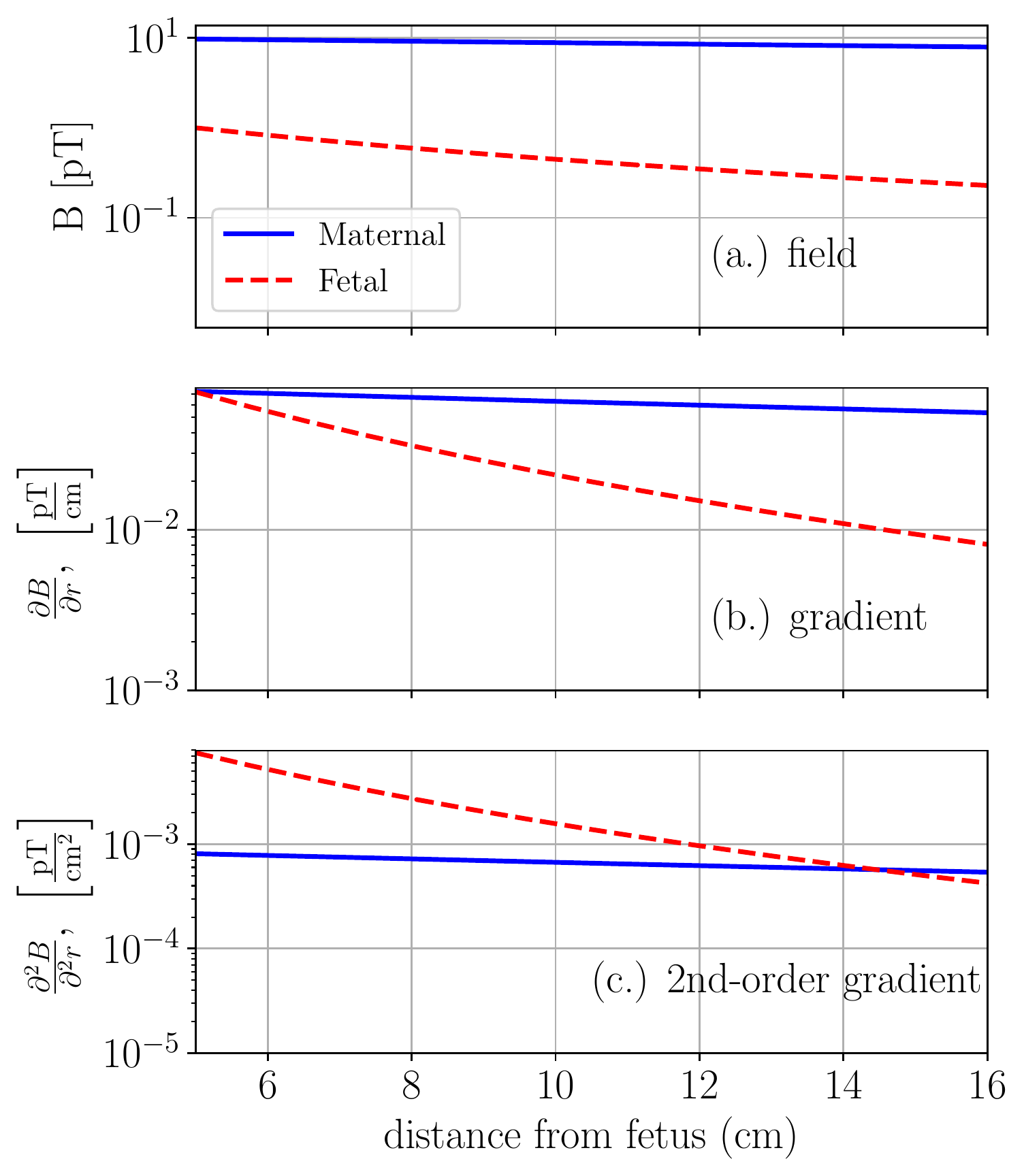}
\caption{Spatial dependence of the fetal and maternal fields with distance. We assume a $ B \propto 1/r^p \, (p=2)$  scaling and plot the magnitude of the field along with its first and second order derivatives. This illustrates the advantage of higher order gradiometry in suppressing the maternal signal. }
\label{fig:rScaling}
\end{figure}

We now consider the question of how effective a gradiometer is in suppressing the maternal background. Modeling the fetal and maternal MCG sources as current dipoles with $B \sim B_0/r^p$, with $p \sim 2$,  we plot in Figure \ref{fig:rScaling} the variation of the fields, their gradients, and their second order gradients with distance from the fetus.  We have assumed that the maternal heart is $r_m = 60$ cm away from the sensor, which is $r_f = 5$ cm from the fetal heart. 

The advantage of using a gradiometer to suppress the maternal contribution to the signal is clear from Figure \ref{fig:rScaling}. We see that at $\sim 5 \,\rm{cm}$ from the fetus,the maternal field is $\sim 10 \times$ larger than the fetal field. The gradients of the field are however comparable in magnitude. A gradient measurement will therefore have less maternal MCG interference. We also see in Figure \ref{fig:rScaling} that we can gain an even greater advantage by detecting the second order gradient. This gain can be obtained if the gradiometers have sufficient sensitivity. For example, in order to detect the fetal gradient at 10 cm with SNR = 1 in 1 second, we surmise (from Figure \ref{fig:rScaling}) that we need a sensor with gradient sensitivity  $\sim 10 \,\rm{fT}\,\rm{cm^{-1} Hz^{-1/2}}$ or better. Likewise, for a second order gradiometer we need a  sensitivity of $ \sim 10 \,\rm{fT}\,\rm{cm^{-2} Hz^{-1/2}}$

Another way to quantify the advantage of the gradiometer is to consider Equation \ref{eq:FOM1}. The idea here is that the background to be suppressed is the maternal field $B_m \sim B_0/r^p$. Assuming that the magnetometer has baseline $L$, we then have that ${\delta B_u = B_f L p/r_f^{p+1}}$, the gradient of the fetal field $\times$ the baseline, and that $\delta B_c= B_m/r_m^p$, the average value of the maternal field at the gradiometer. We substitute those values into equation \ref{eq:FOM1} and plot $\mathcal{F}$ for different baselines and CMRRs. 

\begin{figure}[h!]
	\centering
		\includegraphics[width = \linewidth]{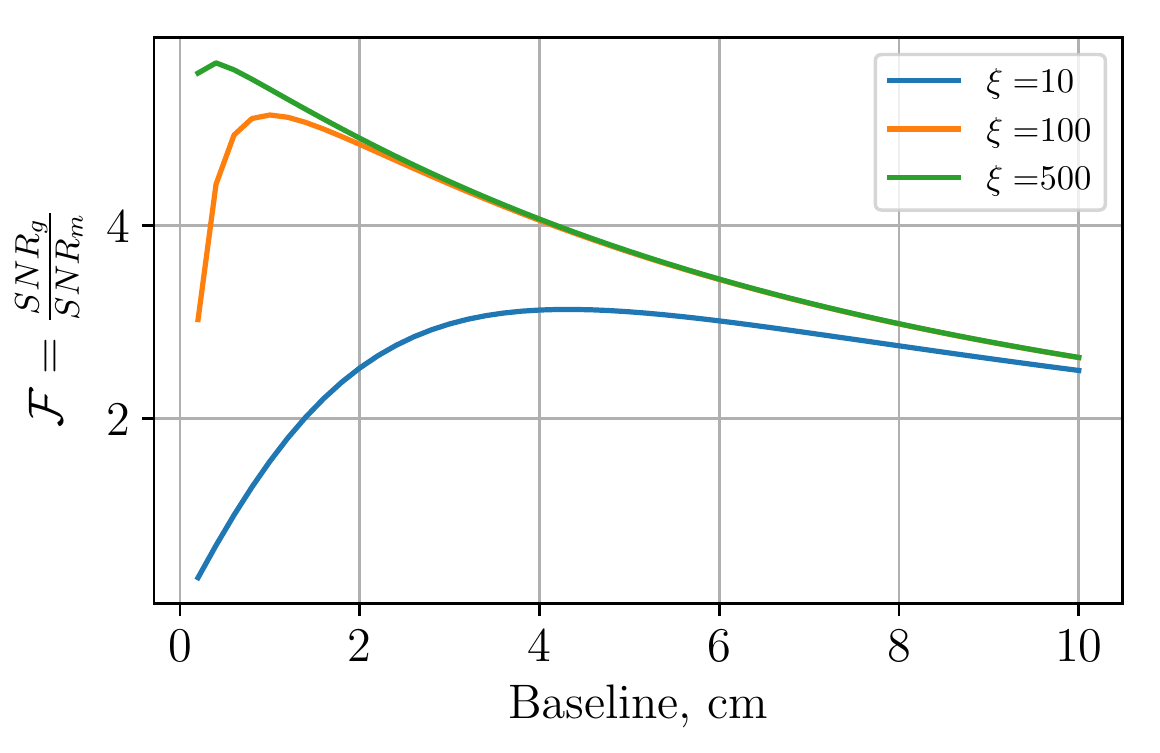}
\caption{Figure of merit for detecting a fetal field valuated for the case in which the dominant background is the maternal field as shown in figure \ref{fig:ECGSYN}. This calculation illustrates that for our operating condition, $\mathcal{F}$ is limited by the magnitude of the gradient of the maternal field for $\xi > 100$}
\label{fig:FOMtrend}
\end{figure}

The point to note from Figure \ref{fig:FOMtrend} is that using a gradiometer  for $\xi >100$, the figure of merit is limited by the maternal field gradient, and not the CMRR. With regards to suppressing the maternal signal therefore, there is no advantage to working to increase the CMRR; time is better spent developing a higher order gradiometer. Furthermore, baselines on the order of 1cm are best suited for maternal MCG suppression.

\section{Comparison to Previous Work}

A number of different AM gradiometer implementations have been reported  in the literature\cite{Affolderbach,Dang,Kominis,Kim,Kamada,Sheng,Sheng2017,Johnson,Shoujun}. Baselines ranging from 0.3 to 10 cm have been attained by either (a) imaging a probe laser going through a single cell onto a segmented photo-detector such as by Kominis et al. \cite{Kominis}, (b) using spatially separated sensors which requires having multiple probe beams, as in this work, or (c) having a single beam going through multiple cells sequentially as done by Kamada et al. \cite{Kamada}. Some features of the design choices are that in case (a) the baseline is limited to the size of the cell $\sim 1 \,\mathrm{cm}$. One advantage it offers is that atomic responses from the two spatial regions comprising the gradiometer may be very similar as the atoms occupy nearly the same volume. This will enable high CMRR. The designs in cases (b) and (c) allow for larger baselines. For case (c), the subtraction of the signal is effectively performed by Faraday rotation of the probe and not by an electronic circuit or in software. As a result, photon shot noise factors in the total noise budget only once, unlike in case (b) where it factors in twice. 

\begin{table}[h!]

	\begin{tabular}{l|c|c|c|c|c|c}
	\hline 
Type   	& Baseline, L               	&Mag. Sens                                  &$\xi$             &  $\mathcal{F}$ &   bw   & Ref                  \\
		& (cm)	                        & $(\rm{fT/\sqrt{Hz}})$                     &      	            &    	    &    (Hz)    &                \\
      \hline
      \hline
    SQUID       & 5                                 & 5 &100                        & 60                &  $1-150$             & \footnotemark[1]\\
    SERF        & 4                                 & 10 & 350                      & 184               & $1-150$               &\footnotemark[2]\\ 
    SERF        & 2                                 & 10 & 750\footnotemark[4]                      & 243               & $1-150$              &\cite{Sheng2017} \\  
    SERF        & 0.8                               & 14 & 40                        &                  & $40-400$             &   \cite{Jiang2018}\\  
    NMOR        & 2.5                               & 80 & $>20$                    &                &               &\cite{Shoujun}\\
    SERF        & 3                                 & 9.3 & $>6$                    &               &               &\cite{Kamada}\\ 
      SERF      & 0.5                               &  .16& $<5$\footnotemark[4]    &                   &               &                      \cite{Dang}\\
      SERF      & $0.3 -  0.9$ \footnotemark[3]     & .54 & $> 13$\footnotemark[4]     &                   &               &                      \cite{Kominis}\\
      SERF      & $0.75 - 12$ \footnotemark[3]      & 4 & $> 17$\footnotemark[4]        &                   &               &    \cite{Kim}\\
      SERF      & 0.5                               & 5 & $> 3$\footnotemark[4]     &                   &               &    \cite{Johnson}\\
      SERF      & $> 2.5$                           & 5 &$> 6$\footnotemark[4]      &                   &               &    \cite{Fang}\\
      CPT       & 1.5                               & 2600 & $> 800$             &                   &               &                       \cite{Affolderbach}\\
      CsOPM     & 10                                & 54 & $> 5$\footnotemark[4]    &                   &               &    \cite{Koch}\\
      Mx        & 5                                 & 300 & 1000                    &                   &               &   \cite{Bison}\\
      \hline

	\end{tabular}
    \footnotetext[1]{Specifications for Tristan 624 Bio-magnetometer from Tristan Technologies, San Diego. Operational bandwidth can exceed a kHz. Only 1- 150 Hz is considered in this work to make comparisons with AM gradiometers} 
    \footnotetext[2]{This Work}
	\footnotetext[3]{Variable baseline: array of sensors. }
	\footnotetext[4]{Estimated from published data}

       \caption{Summary of published results of atomic gradiometers. The figure of merit $\mathcal{F}$ is calculated for $r_s =$ 5 cm, and $p = 2$--typical values for fMCG \cite{Cohen1979}. $\mathcal{F}$ is calculated for an ambient environment sufficiently noisy such that $\left(\delta B_u / \delta B_c \right)^2 \ll \xi^{-2}$ and is reported for sensors with well measured $\xi$ and favorable bandwidth for fMCG measurements.  The values of $\xi$ for previously-published work should be considered lower bounds.}
   \label{tab:FOMtable}

\end{table}

We compare the different AM gradiometer implementations in Table \ref{tab:FOMtable}, listing their baselines and the magnetometric sensitivities. We also list  CMRRs in situations where it is carefully measured by applying large amplitude noise signal and measuring its suppression. In cases where the CMRR is not well measured, we report the lower limit. Given that the sensitivities and CMRR are functions of frequency, we report their average values over a frequency band of interest. For fMCG, that band is roughly 1 - 100 Hz.  Addressing the question of which implementation is best suited for fMCG, we evaluate the figure of merit $\mathcal{F}$ for a source distance characteristic of fMCG in this frequency band.

The evaluated figure of merit $\mathcal{F}$ in Table \ref{tab:FOMtable} can be used to assess the suitability of the gradiometer for fMCG in a noisy environment. It does not quantify the degree to which the maternal MCG signal can be suppressed in the fMCG measurement. As discussed in section \ref{sec:MCGsep}, that limitation is due to the non-negligible gradient of the maternal MCG field that is not suppressed by a first order gradiometer.

We note from Table \ref{tab:FOMtable} that SERF sensors can achieve comparable and even superior sensitivity and CMRR to SQUIDs. The achieved SERF sensitivity is well established. We have shown here that the CMRR can reliably operate at a few $\times 10^2$ over a large frequency range while maintaining this sensitivity. 

The chip scale SERF gradiometers described by Sheng et al. \cite{Sheng2017} have the largest figure of merit for an fMCG application of all the sensors we've considered. In fact, they report that for the cost of a reduced bandwidth, their closed loop gradiometer can achieve a CMRR of up to 1000 at 10Hz. This is very promising for biomagnetism and low field NMR studies\cite{Jiang2018}. 

Compared to the chip scale sensors, magnetometers like ours with volumes $\sim \, \rm{cm^3}$ have an advantage of a a fundamentally lower spin projection noise level. That sensitivity can be exploited once other sources of technical and uncorrelated magnetic noise are reduced..

As a final point, we recommend that subsequent studies in the development of atomic magnetic gradiometers should report the frequency dependence of their CMRR -- measured by adding noise in the frequency range of interest. This specification will enable a more meaningful comparison of techniques and suitability for applications.

\section{Conclusion}

One key  result from this work is that due to differences in temperature, laser intensity distributions, cell wall conditions etc, the individual AM dynamic responses can be different to a degree in which the CMRR can be compromised. As such, we expect that superior gradiometry performance will be obtained if, as described above, a calibration procedure which allows for individual channel differences -- factoring their full complex responses -- is applied to the sensors comprising the gradiometers. We suspect that even implementations with short baselines where atoms occupy nearly the same volume stand to benefit from this.

We have also in this work introduced a figure of merit $\mathcal{F}$ derived from the gradiometer CMRR, its baseline, and the geometric scaling of the signal and noise background. Knowing this figure of merit will help users judge the optimality of a sensor for a particular application. 
If the noise background is from a well-known sources (e.g. a current dipole), the noise background scaling can be described analytically. Otherwise, it can be estimated by measuring the gradient noise for a number of baseline values. The optimal baseline is the one that maximizes $\mathcal{F}$

We have demonstrated a gradiometer constructed from two atomic magnetometers with baseline $L=4$ and common-mode rejection ratio $\sim150$ in the DC to 100 Hz band operating in closed loop. Its  performance is comparable with current commercially-available SQUIDs. Future work will include efforts to combine the benefits of parametric-modulation with active feedback to create a superior gradiometer device.

Finally, we note that for the fMCG detection problem, higher order gradiometry is needed. The gradient in the field arising from the maternal MCG which is not suppressed by a first order gradiometer can be comparable to the fMCG. The development of larger count AM gradiometer arrays should therefore be prioritized.

This work was supported by the NIH Eunice Kennedy Shriver National Institute of Child Health \& Human Development, No. R01HD057965.

\bibliographystyle{aipnum4-1}
\bibliography{GradioRefs}

\begin{thebibliography}{34}%
\makeatletter
\providecommand \@ifxundefined [1]{%
 \@ifx{#1\undefined}
}%
\providecommand \@ifnum [1]{%
 \ifnum #1\expandafter \@firstoftwo
 \else \expandafter \@secondoftwo
 \fi
}%
\providecommand \@ifx [1]{%
 \ifx #1\expandafter \@firstoftwo
 \else \expandafter \@secondoftwo
 \fi
}%
\providecommand \natexlab [1]{#1}%
\providecommand \enquote  [1]{``#1''}%
\providecommand \bibnamefont  [1]{#1}%
\providecommand \bibfnamefont [1]{#1}%
\providecommand \citenamefont [1]{#1}%
\providecommand \href@noop [0]{\@secondoftwo}%
\providecommand \href [0]{\begingroup \@sanitize@url \@href}%
\providecommand \@href[1]{\@@startlink{#1}\@@href}%
\providecommand \@@href[1]{\endgroup#1\@@endlink}%
\providecommand \@sanitize@url [0]{\catcode `\\12\catcode `\$12\catcode
  `\&12\catcode `\#12\catcode `\^12\catcode `\_12\catcode `\%12\relax}%
\providecommand \@@startlink[1]{}%
\providecommand \@@endlink[0]{}%
\providecommand \url  [0]{\begingroup\@sanitize@url \@url }%
\providecommand \@url [1]{\endgroup\@href {#1}{\urlprefix }}%
\providecommand \urlprefix  [0]{URL }%
\providecommand \Eprint [0]{\href }%
\providecommand \doibase [0]{http://dx.doi.org/}%
\providecommand \selectlanguage [0]{\@gobble}%
\providecommand \bibinfo  [0]{\@secondoftwo}%
\providecommand \bibfield  [0]{\@secondoftwo}%
\providecommand \translation [1]{[#1]}%
\providecommand \BibitemOpen [0]{}%
\providecommand \bibitemStop [0]{}%
\providecommand \bibitemNoStop [0]{.\EOS\space}%
\providecommand \EOS [0]{\spacefactor3000\relax}%
\providecommand \BibitemShut  [1]{\csname bibitem#1\endcsname}%
\let\auto@bib@innerbib\@empty
\bibitem [{\citenamefont {Colombo}\ \emph {et~al.}(2016)\citenamefont
  {Colombo}, \citenamefont {Carter}, \citenamefont {Borna}, \citenamefont
  {Jau}, \citenamefont {Johnson}, \citenamefont {Dagel},\ and\ \citenamefont
  {Schwindt}}]{Colombo2016}%
  \BibitemOpen
  \bibfield  {author} {\bibinfo {author} {\bibfnamefont {A.~P.}\ \bibnamefont
  {Colombo}}, \bibinfo {author} {\bibfnamefont {T.~R.}\ \bibnamefont {Carter}},
  \bibinfo {author} {\bibfnamefont {A.}~\bibnamefont {Borna}}, \bibinfo
  {author} {\bibfnamefont {Y.-Y.}\ \bibnamefont {Jau}}, \bibinfo {author}
  {\bibfnamefont {C.~N.}\ \bibnamefont {Johnson}}, \bibinfo {author}
  {\bibfnamefont {A.~L.}\ \bibnamefont {Dagel}}, \ and\ \bibinfo {author}
  {\bibfnamefont {P.~D.~D.}\ \bibnamefont {Schwindt}},\ }\href {\doibase
  10.1364/OE.24.015403} {\bibfield  {journal} {\bibinfo  {journal} {Opt.
  Express}\ }\textbf {\bibinfo {volume} {24}},\ \bibinfo {pages} {15403}
  (\bibinfo {year} {2016})}\BibitemShut {NoStop}%
\bibitem [{\citenamefont {Sander}\ \emph {et~al.}(2012)\citenamefont {Sander},
  \citenamefont {Preusser}, \citenamefont {Mhaskar}, \citenamefont {Kitching},
  \citenamefont {Trahms},\ and\ \citenamefont {Knappe}}]{Sander2012}%
  \BibitemOpen
  \bibfield  {author} {\bibinfo {author} {\bibfnamefont {T.~H.}\ \bibnamefont
  {Sander}}, \bibinfo {author} {\bibfnamefont {J.}~\bibnamefont {Preusser}},
  \bibinfo {author} {\bibfnamefont {R.}~\bibnamefont {Mhaskar}}, \bibinfo
  {author} {\bibfnamefont {J.}~\bibnamefont {Kitching}}, \bibinfo {author}
  {\bibfnamefont {L.}~\bibnamefont {Trahms}}, \ and\ \bibinfo {author}
  {\bibfnamefont {S.}~\bibnamefont {Knappe}},\ }\href {\doibase
  10.1364/BOE.3.000981} {\bibfield  {journal} {\bibinfo  {journal} {Biomed.
  Opt. Express}\ }\textbf {\bibinfo {volume} {3}},\ \bibinfo {pages} {981}
  (\bibinfo {year} {2012})}\BibitemShut {NoStop}%
\bibitem [{\citenamefont {Keenan}, \citenamefont {Blay},\ and\ \citenamefont
  {Romans}(2011)}]{Keenan2011}%
  \BibitemOpen
  \bibfield  {author} {\bibinfo {author} {\bibfnamefont {S.~T.}\ \bibnamefont
  {Keenan}}, \bibinfo {author} {\bibfnamefont {K.~R.}\ \bibnamefont {Blay}}, \
  and\ \bibinfo {author} {\bibfnamefont {E.~J.}\ \bibnamefont {Romans}},\
  }\href {http://stacks.iop.org/0953-2048/24/i=8/a=085019} {\bibfield
  {journal} {\bibinfo  {journal} {Superconductor Science and Technology}\
  }\textbf {\bibinfo {volume} {24}},\ \bibinfo {pages} {085019} (\bibinfo
  {year} {2011})}\BibitemShut {NoStop}%
\bibitem [{\citenamefont {{Wiegert}}\ and\ \citenamefont
  {{Oeschger}}(2006)}]{ordnance2}%
  \BibitemOpen
  \bibfield  {author} {\bibinfo {author} {\bibfnamefont {R.}~\bibnamefont
  {{Wiegert}}}\ and\ \bibinfo {author} {\bibfnamefont {J.}~\bibnamefont
  {{Oeschger}}},\ }in\ \href {\doibase 10.1109/OCEANS.2006.306805} {\emph
  {\bibinfo {booktitle} {OCEANS 2006}}}\ (\bibinfo {year} {2006})\ pp.\
  \bibinfo {pages} {1--6}\BibitemShut {NoStop}%
\bibitem [{\citenamefont {Jiang}\ \emph {et~al.}(2019)\citenamefont {Jiang},
  \citenamefont {Frutos}, \citenamefont {Wu}, \citenamefont {Blanchard},
  \citenamefont {Peng},\ and\ \citenamefont {Budker}}]{Jiang2018}%
  \BibitemOpen
  \bibfield  {author} {\bibinfo {author} {\bibfnamefont {M.}~\bibnamefont
  {Jiang}}, \bibinfo {author} {\bibfnamefont {R.~P.}\ \bibnamefont {Frutos}},
  \bibinfo {author} {\bibfnamefont {T.}~\bibnamefont {Wu}}, \bibinfo {author}
  {\bibfnamefont {J.~W.}\ \bibnamefont {Blanchard}}, \bibinfo {author}
  {\bibfnamefont {X.}~\bibnamefont {Peng}}, \ and\ \bibinfo {author}
  {\bibfnamefont {D.}~\bibnamefont {Budker}},\ }\href {\doibase
  10.1103/PhysRevApplied.11.024005} {\bibfield  {journal} {\bibinfo  {journal}
  {Phys. Rev. Applied}\ }\textbf {\bibinfo {volume} {11}},\ \bibinfo {pages}
  {024005} (\bibinfo {year} {2019})}\BibitemShut {NoStop}%
\bibitem [{\citenamefont {Quartero}\ \emph {et~al.}(2002)\citenamefont
  {Quartero}, \citenamefont {Stinstra}, \citenamefont {Golbach}, \citenamefont
  {Meijboom},\ and\ \citenamefont {Peters}}]{Quartero2002}%
  \BibitemOpen
  \bibfield  {author} {\bibinfo {author} {\bibfnamefont {H.~W.~P.}\
  \bibnamefont {Quartero}}, \bibinfo {author} {\bibfnamefont {J.~G.}\
  \bibnamefont {Stinstra}}, \bibinfo {author} {\bibfnamefont {E.~G.~M.}\
  \bibnamefont {Golbach}}, \bibinfo {author} {\bibfnamefont {E.~J.}\
  \bibnamefont {Meijboom}}, \ and\ \bibinfo {author} {\bibfnamefont {M.~J.}\
  \bibnamefont {Peters}},\ }\href {\doibase 10.1046/j.1469-0705.2002.00754.x}
  {\bibfield  {journal} {\bibinfo  {journal} {Ultrasound in Obstetrics and
  Gynecology}\ }\textbf {\bibinfo {volume} {20}},\ \bibinfo {pages} {142}
  (\bibinfo {year} {2002})}\BibitemShut {NoStop}%
\bibitem [{\citenamefont {Strasburger}, \citenamefont {Cheulkar},\ and\
  \citenamefont {Wakai}(2008)}]{Strasburger2008}%
  \BibitemOpen
  \bibfield  {author} {\bibinfo {author} {\bibfnamefont {J.~F.}\ \bibnamefont
  {Strasburger}}, \bibinfo {author} {\bibfnamefont {B.}~\bibnamefont
  {Cheulkar}}, \ and\ \bibinfo {author} {\bibfnamefont {R.~T.}\ \bibnamefont
  {Wakai}},\ }\href {\doibase 10.1016/j.hrthm.2008.02.035} {\bibfield
  {journal} {\bibinfo  {journal} {Heart Rhythm}\ }\textbf {\bibinfo {volume}
  {5}},\ \bibinfo {pages} {1073} (\bibinfo {year} {2008})}\BibitemShut
  {NoStop}%
\bibitem [{\citenamefont {H\"am\"al\"ainen}\ \emph {et~al.}(1993)\citenamefont
  {H\"am\"al\"ainen}, \citenamefont {Hari}, \citenamefont {Ilmoniemi},
  \citenamefont {Knuutila},\ and\ \citenamefont {Lounasmaa}}]{Finland1993}%
  \BibitemOpen
  \bibfield  {author} {\bibinfo {author} {\bibfnamefont {M.}~\bibnamefont
  {H\"am\"al\"ainen}}, \bibinfo {author} {\bibfnamefont {R.}~\bibnamefont
  {Hari}}, \bibinfo {author} {\bibfnamefont {R.~J.}\ \bibnamefont {Ilmoniemi}},
  \bibinfo {author} {\bibfnamefont {J.}~\bibnamefont {Knuutila}}, \ and\
  \bibinfo {author} {\bibfnamefont {O.~V.}\ \bibnamefont {Lounasmaa}},\ }\href
  {\doibase 10.1103/RevModPhys.65.413} {\bibfield  {journal} {\bibinfo
  {journal} {Rev. Mod. Phys.}\ }\textbf {\bibinfo {volume} {65}},\ \bibinfo
  {pages} {413} (\bibinfo {year} {1993})}\BibitemShut {NoStop}%
\bibitem [{\citenamefont {Clarke}\ and\ \citenamefont
  {Braginski}(2009)}]{Clarke2004}%
  \BibitemOpen
  \bibfield  {author} {\bibinfo {author} {\bibfnamefont {J.}~\bibnamefont
  {Clarke}}\ and\ \bibinfo {author} {\bibfnamefont {A.~I.}\ \bibnamefont
  {Braginski}},\ }\href@noop {} {\emph {\bibinfo {title} {The SQUID
  Handbook}}}\ (\bibinfo  {publisher} {John Wiley and Sons, Inc},\ \bibinfo
  {year} {2009})\BibitemShut {NoStop}%
\bibitem [{\citenamefont {Affolderbach}\ \emph {et~al.}(2002)\citenamefont
  {Affolderbach}, \citenamefont {Stahler}, \citenamefont {Knappe},\ and\
  \citenamefont {Wynands}}]{Affolderbach}%
  \BibitemOpen
  \bibfield  {author} {\bibinfo {author} {\bibfnamefont {C.}~\bibnamefont
  {Affolderbach}}, \bibinfo {author} {\bibfnamefont {M.}~\bibnamefont
  {Stahler}}, \bibinfo {author} {\bibfnamefont {S.}~\bibnamefont {Knappe}}, \
  and\ \bibinfo {author} {\bibfnamefont {R.}~\bibnamefont {Wynands}},\ }\href
  {\doibase 10.1007/s00340-002-0959-8} {\bibfield  {journal} {\bibinfo
  {journal} {Applied Physics B}\ }\textbf {\bibinfo {volume} {75}},\ \bibinfo
  {pages} {605} (\bibinfo {year} {2002})}\BibitemShut {NoStop}%
\bibitem [{\citenamefont {Dang}, \citenamefont {Maloof},\ and\ \citenamefont
  {Romalis}(2010)}]{Dang}%
  \BibitemOpen
  \bibfield  {author} {\bibinfo {author} {\bibfnamefont {H.~B.}\ \bibnamefont
  {Dang}}, \bibinfo {author} {\bibfnamefont {A.~C.}\ \bibnamefont {Maloof}}, \
  and\ \bibinfo {author} {\bibfnamefont {M.~V.}\ \bibnamefont {Romalis}},\
  }\href {\doibase http://dx.doi.org/10.1063/1.3491215} {\bibfield  {journal}
  {\bibinfo  {journal} {Applied Physics Letters}\ }\textbf {\bibinfo {volume}
  {97}},\ \bibinfo {eid} {151110} (\bibinfo {year} {2010})}\BibitemShut
  {NoStop}%
\bibitem [{\citenamefont {Kominis}\ \emph {et~al.}(2003)\citenamefont
  {Kominis}, \citenamefont {Kornack}, \citenamefont {Allred},\ and\
  \citenamefont {Romalis}}]{Kominis}%
  \BibitemOpen
  \bibfield  {author} {\bibinfo {author} {\bibfnamefont {I.~K.}\ \bibnamefont
  {Kominis}}, \bibinfo {author} {\bibfnamefont {T.~W.}\ \bibnamefont
  {Kornack}}, \bibinfo {author} {\bibfnamefont {J.~C.}\ \bibnamefont {Allred}},
  \ and\ \bibinfo {author} {\bibfnamefont {M.~V.}\ \bibnamefont {Romalis}},\
  }\href {\doibase http://dx.doi.org/10.1038/nature01484} {\bibfield  {journal}
  {\bibinfo  {journal} {Nature}\ }\textbf {\bibinfo {volume} {422}},\ \bibinfo
  {pages} {596} (\bibinfo {year} {2003})}\BibitemShut {NoStop}%
\bibitem [{\citenamefont {Kim}\ \emph {et~al.}(2014)\citenamefont {Kim},
  \citenamefont {Begus}, \citenamefont {Xia}, \citenamefont {Lee},
  \citenamefont {Jazbinsek}, \citenamefont {Trontelj},\ and\ \citenamefont
  {Romalis}}]{Kim}%
  \BibitemOpen
  \bibfield  {author} {\bibinfo {author} {\bibfnamefont {K.}~\bibnamefont
  {Kim}}, \bibinfo {author} {\bibfnamefont {S.}~\bibnamefont {Begus}}, \bibinfo
  {author} {\bibfnamefont {H.}~\bibnamefont {Xia}}, \bibinfo {author}
  {\bibfnamefont {S.-K.}\ \bibnamefont {Lee}}, \bibinfo {author} {\bibfnamefont
  {V.}~\bibnamefont {Jazbinsek}}, \bibinfo {author} {\bibfnamefont
  {Z.}~\bibnamefont {Trontelj}}, \ and\ \bibinfo {author} {\bibfnamefont
  {M.~V.}\ \bibnamefont {Romalis}},\ }\href {\doibase
  http://dx.doi.org/10.1016/j.neuroimage.2013.10.040} {\bibfield  {journal}
  {\bibinfo  {journal} {NeuroImage}\ }\textbf {\bibinfo {volume} {89}},\
  \bibinfo {pages} {143 } (\bibinfo {year} {2014})}\BibitemShut {NoStop}%
\bibitem [{\citenamefont {Kamada}\ \emph {et~al.}(2015)\citenamefont {Kamada},
  \citenamefont {Ito}, \citenamefont {Ichihara}, \citenamefont {Mizutani},\
  and\ \citenamefont {Kobayashi}}]{Kamada}%
  \BibitemOpen
  \bibfield  {author} {\bibinfo {author} {\bibfnamefont {K.}~\bibnamefont
  {Kamada}}, \bibinfo {author} {\bibfnamefont {Y.}~\bibnamefont {Ito}},
  \bibinfo {author} {\bibfnamefont {S.}~\bibnamefont {Ichihara}}, \bibinfo
  {author} {\bibfnamefont {N.}~\bibnamefont {Mizutani}}, \ and\ \bibinfo
  {author} {\bibfnamefont {T.}~\bibnamefont {Kobayashi}},\ }\href {\doibase
  10.1364/OE.23.006976} {\bibfield  {journal} {\bibinfo  {journal} {Opt.
  Express}\ }\textbf {\bibinfo {volume} {23}},\ \bibinfo {pages} {6976}
  (\bibinfo {year} {2015})}\BibitemShut {NoStop}%
\bibitem [{\citenamefont {Sheng}\ \emph {et~al.}(2013)\citenamefont {Sheng},
  \citenamefont {Li}, \citenamefont {Dural},\ and\ \citenamefont
  {Romalis}}]{Sheng}%
  \BibitemOpen
  \bibfield  {author} {\bibinfo {author} {\bibfnamefont {D.}~\bibnamefont
  {Sheng}}, \bibinfo {author} {\bibfnamefont {S.}~\bibnamefont {Li}}, \bibinfo
  {author} {\bibfnamefont {N.}~\bibnamefont {Dural}}, \ and\ \bibinfo {author}
  {\bibfnamefont {M.~V.}\ \bibnamefont {Romalis}},\ }\href {\doibase
  10.1103/PhysRevLett.110.160802} {\bibfield  {journal} {\bibinfo  {journal}
  {Phys. Rev. Lett.}\ }\textbf {\bibinfo {volume} {110}},\ \bibinfo {pages}
  {160802} (\bibinfo {year} {2013})}\BibitemShut {NoStop}%
\bibitem [{\citenamefont {Sheng}\ \emph {et~al.}(2017)\citenamefont {Sheng},
  \citenamefont {Perry}, \citenamefont {Krzyzewski}, \citenamefont {Geller},
  \citenamefont {Kitching},\ and\ \citenamefont {Knappe}}]{Sheng2017}%
  \BibitemOpen
  \bibfield  {author} {\bibinfo {author} {\bibfnamefont {D.}~\bibnamefont
  {Sheng}}, \bibinfo {author} {\bibfnamefont {A.~R.}\ \bibnamefont {Perry}},
  \bibinfo {author} {\bibfnamefont {S.~P.}\ \bibnamefont {Krzyzewski}},
  \bibinfo {author} {\bibfnamefont {S.}~\bibnamefont {Geller}}, \bibinfo
  {author} {\bibfnamefont {J.}~\bibnamefont {Kitching}}, \ and\ \bibinfo
  {author} {\bibfnamefont {S.}~\bibnamefont {Knappe}},\ }\href {\doibase
  10.1063/1.4974349} {\bibfield  {journal} {\bibinfo  {journal} {Applied
  Physics Letters}\ }\textbf {\bibinfo {volume} {110}},\ \bibinfo {pages}
  {031106} (\bibinfo {year} {2017})}\BibitemShut {NoStop}%
\bibitem [{\citenamefont {Johnson}, \citenamefont {Schwindt},\ and\
  \citenamefont {Weisend}(2010)}]{Johnson}%
  \BibitemOpen
  \bibfield  {author} {\bibinfo {author} {\bibfnamefont {C.}~\bibnamefont
  {Johnson}}, \bibinfo {author} {\bibfnamefont {P.~D.~D.}\ \bibnamefont
  {Schwindt}}, \ and\ \bibinfo {author} {\bibfnamefont {M.}~\bibnamefont
  {Weisend}},\ }\href {\doibase http://dx.doi.org/10.1063/1.3522648} {\bibfield
   {journal} {\bibinfo  {journal} {Applied Physics Letters}\ }\textbf {\bibinfo
  {volume} {97}},\ \bibinfo {eid} {243703} (\bibinfo {year}
  {2010})}\BibitemShut {NoStop}%
\bibitem [{\citenamefont {Xu}\ \emph {et~al.}(2006)\citenamefont {Xu},
  \citenamefont {Rochester}, \citenamefont {Yashchuk}, \citenamefont
  {Donaldson},\ and\ \citenamefont {Budker}}]{Shoujun}%
  \BibitemOpen
  \bibfield  {author} {\bibinfo {author} {\bibfnamefont {S.}~\bibnamefont
  {Xu}}, \bibinfo {author} {\bibfnamefont {S.~M.}\ \bibnamefont {Rochester}},
  \bibinfo {author} {\bibfnamefont {V.~V.}\ \bibnamefont {Yashchuk}}, \bibinfo
  {author} {\bibfnamefont {M.~H.}\ \bibnamefont {Donaldson}}, \ and\ \bibinfo
  {author} {\bibfnamefont {D.}~\bibnamefont {Budker}},\ }\href {\doibase
  http://dx.doi.org/10.1063/1.2336087} {\bibfield  {journal} {\bibinfo
  {journal} {Review of Scientific Instruments}\ }\textbf {\bibinfo {volume}
  {77}},\ \bibinfo {eid} {083106} (\bibinfo {year} {2006})}\BibitemShut
  {NoStop}%
\bibitem [{\citenamefont {Wyllie}\ \emph {et~al.}(2012)\citenamefont {Wyllie},
  \citenamefont {Kauer}, \citenamefont {Smetana}, \citenamefont {Wakai},\ and\
  \citenamefont {Walker}}]{Wyllie}%
  \BibitemOpen
  \bibfield  {author} {\bibinfo {author} {\bibfnamefont {R.}~\bibnamefont
  {Wyllie}}, \bibinfo {author} {\bibfnamefont {M.}~\bibnamefont {Kauer}},
  \bibinfo {author} {\bibfnamefont {G.~S.}\ \bibnamefont {Smetana}}, \bibinfo
  {author} {\bibfnamefont {R.~T.}\ \bibnamefont {Wakai}}, \ and\ \bibinfo
  {author} {\bibfnamefont {T.~G.}\ \bibnamefont {Walker}},\ }\href
  {http://stacks.iop.org/0031-9155/57/i=9/a=2619} {\bibfield  {journal}
  {\bibinfo  {journal} {Physics in Medicine and Biology}\ }\textbf {\bibinfo
  {volume} {57}},\ \bibinfo {pages} {2619} (\bibinfo {year}
  {2012})}\BibitemShut {NoStop}%
\bibitem [{\citenamefont {Sulai}\ \emph {et~al.}(2013)\citenamefont {Sulai},
  \citenamefont {Wyllie}, \citenamefont {Kauer}, \citenamefont {Smetana},
  \citenamefont {Wakai},\ and\ \citenamefont {Walker}}]{Sulai2013}%
  \BibitemOpen
  \bibfield  {author} {\bibinfo {author} {\bibfnamefont {I.~A.}\ \bibnamefont
  {Sulai}}, \bibinfo {author} {\bibfnamefont {R.}~\bibnamefont {Wyllie}},
  \bibinfo {author} {\bibfnamefont {M.}~\bibnamefont {Kauer}}, \bibinfo
  {author} {\bibfnamefont {G.~S.}\ \bibnamefont {Smetana}}, \bibinfo {author}
  {\bibfnamefont {R.~T.}\ \bibnamefont {Wakai}}, \ and\ \bibinfo {author}
  {\bibfnamefont {T.~G.}\ \bibnamefont {Walker}},\ }\href {\doibase
  10.1364/OL.38.000974} {\bibfield  {journal} {\bibinfo  {journal} {Opt.
  Lett.}\ }\textbf {\bibinfo {volume} {38}},\ \bibinfo {pages} {974} (\bibinfo
  {year} {2013})}\BibitemShut {NoStop}%
\bibitem [{\citenamefont {Allred}\ \emph {et~al.}(2002)\citenamefont {Allred},
  \citenamefont {Lyman}, \citenamefont {Kornack},\ and\ \citenamefont
  {Romalis}}]{Allred}%
  \BibitemOpen
  \bibfield  {author} {\bibinfo {author} {\bibfnamefont {J.~C.}\ \bibnamefont
  {Allred}}, \bibinfo {author} {\bibfnamefont {R.~N.}\ \bibnamefont {Lyman}},
  \bibinfo {author} {\bibfnamefont {T.~W.}\ \bibnamefont {Kornack}}, \ and\
  \bibinfo {author} {\bibfnamefont {M.~V.}\ \bibnamefont {Romalis}},\ }\href
  {\doibase 10.1103/PhysRevLett.89.130801} {\bibfield  {journal} {\bibinfo
  {journal} {Phys. Rev. Lett.}\ }\textbf {\bibinfo {volume} {89}},\ \bibinfo
  {pages} {130801} (\bibinfo {year} {2002})}\BibitemShut {NoStop}%
\bibitem [{\citenamefont {Wyllie~IV}(2012)}]{WyllieThesis}%
  \BibitemOpen
  \bibfield  {author} {\bibinfo {author} {\bibfnamefont {R.}~\bibnamefont
  {Wyllie~IV}},\ }\emph {\bibinfo {title} {The Development of a Multichannel
  Atomic Magnetometer Array for Fetal Magnetocardiography}},\ \href@noop {}
  {Ph.D. thesis},\ \bibinfo  {school} {University of Wisconsin-Madison}
  (\bibinfo {year} {2012})\BibitemShut {NoStop}%
\bibitem [{\citenamefont {Li}, \citenamefont {Wakai},\ and\ \citenamefont
  {Walker}(2006)}]{Li}%
  \BibitemOpen
  \bibfield  {author} {\bibinfo {author} {\bibfnamefont {Z.}~\bibnamefont
  {Li}}, \bibinfo {author} {\bibfnamefont {R.~T.}\ \bibnamefont {Wakai}}, \
  and\ \bibinfo {author} {\bibfnamefont {T.~G.}\ \bibnamefont {Walker}},\
  }\href {\doibase http://dx.doi.org/10.1063/1.2357553} {\bibfield  {journal}
  {\bibinfo  {journal} {Applied Physics Letters}\ }\textbf {\bibinfo {volume}
  {89}},\ \bibinfo {eid} {134105} (\bibinfo {year} {2006})}\BibitemShut
  {NoStop}%
\bibitem [{\citenamefont {Hobbs}(2004)}]{Hobbs}%
  \BibitemOpen
  \bibfield  {author} {\bibinfo {author} {\bibfnamefont {P.~C.~D.}\
  \bibnamefont {Hobbs}},\ }\href@noop {} {\emph {\bibinfo {title} {Building
  Electro-Optical Systems: Making it all Work}}},\ Vol.~\bibinfo {volume} {I}\
  (\bibinfo  {publisher} {Wiley-VCH},\ \bibinfo {year} {2004})\BibitemShut
  {NoStop}%
\bibitem [{\citenamefont {Deland}(2017)}]{DeLandThesis}%
  \BibitemOpen
  \bibfield  {author} {\bibinfo {author} {\bibfnamefont {Z.~J.}\ \bibnamefont
  {Deland}},\ }\emph {\bibinfo {title} {Advances in Fetal Magnetocardiography
  Using SERF atomic magnetometers}},\ \href@noop {} {Ph.D. thesis},\ \bibinfo
  {school} {University of Wisconsin-Madison} (\bibinfo {year}
  {2017})\BibitemShut {NoStop}%
\bibitem [{\citenamefont {Hornberger}\ and\ \citenamefont
  {Collins}(2008)}]{HORNBERGER2008}%
  \BibitemOpen
  \bibfield  {author} {\bibinfo {author} {\bibfnamefont {L.~K.}\ \bibnamefont
  {Hornberger}}\ and\ \bibinfo {author} {\bibfnamefont {K.}~\bibnamefont
  {Collins}},\ }\href {\doibase https://doi.org/10.1016/j.jacc.2007.09.016}
  {\bibfield  {journal} {\bibinfo  {journal} {Journal of the American College
  of Cardiology}\ }\textbf {\bibinfo {volume} {51}},\ \bibinfo {pages} {85 }
  (\bibinfo {year} {2008})}\BibitemShut {NoStop}%
\bibitem [{\citenamefont {Batie}\ \emph {et~al.}(2018)\citenamefont {Batie},
  \citenamefont {Bitant}, \citenamefont {Strasburger}, \citenamefont {Shah},
  \citenamefont {Alem},\ and\ \citenamefont {Wakai}}]{Batie}%
  \BibitemOpen
  \bibfield  {author} {\bibinfo {author} {\bibfnamefont {M.}~\bibnamefont
  {Batie}}, \bibinfo {author} {\bibfnamefont {S.}~\bibnamefont {Bitant}},
  \bibinfo {author} {\bibfnamefont {J.~F.}\ \bibnamefont {Strasburger}},
  \bibinfo {author} {\bibfnamefont {V.}~\bibnamefont {Shah}}, \bibinfo {author}
  {\bibfnamefont {O.}~\bibnamefont {Alem}}, \ and\ \bibinfo {author}
  {\bibfnamefont {R.~T.}\ \bibnamefont {Wakai}},\ }\href@noop {} {\bibfield
  {journal} {\bibinfo  {journal} {JACC Clin Electrophysiol.}\ }\textbf
  {\bibinfo {volume} {4}},\ \bibinfo {pages} {284} (\bibinfo {year}
  {2018})}\BibitemShut {NoStop}%
\bibitem [{\citenamefont {FDA}()}]{FDAdoc}%
  \BibitemOpen
  \bibfield  {author} {\bibinfo {author} {\bibnamefont {FDA}},\ }\href@noop {}
  {\enquote {\bibinfo {title} {005\_revised 510(k) summary statement for model
  621\_624},}\ }\bibinfo {howpublished}
  {\url{https://www.accessdata.fda.gov/cdrh_docs/pdf15/K151135.pdf}}\BibitemShut
  {NoStop}%
\bibitem [{\citenamefont {Cohen}\ and\ \citenamefont
  {Hosaka}(1976)}]{Cohen1979}%
  \BibitemOpen
  \bibfield  {author} {\bibinfo {author} {\bibfnamefont {D.}~\bibnamefont
  {Cohen}}\ and\ \bibinfo {author} {\bibfnamefont {H.}~\bibnamefont {Hosaka}},\
  }\href {\doibase http://dx.doi.org/10.1016/S0022-0736(76)80041-6} {\bibfield
  {journal} {\bibinfo  {journal} {Journal of Electrocardiology}\ }\textbf
  {\bibinfo {volume} {9}},\ \bibinfo {pages} {409 } (\bibinfo {year}
  {1976})}\BibitemShut {NoStop}%
\bibitem [{\citenamefont {McSharry}\ and\ \citenamefont {Clifford}()}]{ECGSYN}%
  \BibitemOpen
  \bibfield  {author} {\bibinfo {author} {\bibfnamefont {P.}~\bibnamefont
  {McSharry}}\ and\ \bibinfo {author} {\bibfnamefont {G.}~\bibnamefont
  {Clifford}},\ }\href@noop {} {\enquote {\bibinfo {title} {Ecgsyn: A realistic
  ecg waveform generator},}\ }\bibinfo {howpublished}
  {\url{https://www.physionet.org/physiotools/ecgsyn/}},\ \bibinfo {note}
  {accessed 2016.01.07}\BibitemShut {NoStop}%
\bibitem [{\citenamefont {Fang}\ \emph {et~al.}(2014)\citenamefont {Fang},
  \citenamefont {Wang}, \citenamefont {Zhang}, \citenamefont {Li},\ and\
  \citenamefont {Zou}}]{Fang}%
  \BibitemOpen
  \bibfield  {author} {\bibinfo {author} {\bibfnamefont {J.}~\bibnamefont
  {Fang}}, \bibinfo {author} {\bibfnamefont {T.}~\bibnamefont {Wang}}, \bibinfo
  {author} {\bibfnamefont {H.}~\bibnamefont {Zhang}}, \bibinfo {author}
  {\bibfnamefont {Y.}~\bibnamefont {Li}}, \ and\ \bibinfo {author}
  {\bibfnamefont {S.}~\bibnamefont {Zou}},\ }\href {\doibase
  http://dx.doi.org/10.1063/1.4902567} {\bibfield  {journal} {\bibinfo
  {journal} {Review of Scientific Instruments}\ }\textbf {\bibinfo {volume}
  {85}},\ \bibinfo {eid} {123104} (\bibinfo {year} {2014}),\
  http://dx.doi.org/10.1063/1.4902567}\BibitemShut {NoStop}%
\bibitem [{\citenamefont {Koch}\ \emph {et~al.}(2015)\citenamefont {Koch},
  \citenamefont {Bison}, \citenamefont {Grujić}, \citenamefont {Heil},
  \citenamefont {Kasprzak}, \citenamefont {Knowles}, \citenamefont {Kraft},
  \citenamefont {Pazgalev}, \citenamefont {Schnabel}, \citenamefont {Voigt},\
  and\ \citenamefont {Weis}}]{Koch}%
  \BibitemOpen
  \bibfield  {author} {\bibinfo {author} {\bibfnamefont {H.-C.}\ \bibnamefont
  {Koch}}, \bibinfo {author} {\bibfnamefont {G.}~\bibnamefont {Bison}},
  \bibinfo {author} {\bibfnamefont {Z.}~\bibnamefont {Grujić}}, \bibinfo
  {author} {\bibfnamefont {W.}~\bibnamefont {Heil}}, \bibinfo {author}
  {\bibfnamefont {M.}~\bibnamefont {Kasprzak}}, \bibinfo {author}
  {\bibfnamefont {P.}~\bibnamefont {Knowles}}, \bibinfo {author} {\bibfnamefont
  {A.}~\bibnamefont {Kraft}}, \bibinfo {author} {\bibfnamefont
  {A.}~\bibnamefont {Pazgalev}}, \bibinfo {author} {\bibfnamefont
  {A.}~\bibnamefont {Schnabel}}, \bibinfo {author} {\bibfnamefont
  {J.}~\bibnamefont {Voigt}}, \ and\ \bibinfo {author} {\bibfnamefont
  {A.}~\bibnamefont {Weis}},\ }\href {\doibase 10.1140/epjd/e2015-60018-7}
  {\bibfield  {journal} {\bibinfo  {journal} {The European Physical Journal D}\
  }\textbf {\bibinfo {volume} {69}},\ \bibinfo {eid} {202} (\bibinfo {year}
  {2015}),\ 10.1140/epjd/e2015-60018-7}\BibitemShut {NoStop}%
\bibitem [{\citenamefont {Bison}\ \emph {et~al.}(2009)\citenamefont {Bison},
  \citenamefont {Castagna}, \citenamefont {Hofer}, \citenamefont {Knowles},
  \citenamefont {Schenker}, \citenamefont {Kasprzak}, \citenamefont {Saudan},\
  and\ \citenamefont {Weis}}]{Bison}%
  \BibitemOpen
  \bibfield  {author} {\bibinfo {author} {\bibfnamefont {G.}~\bibnamefont
  {Bison}}, \bibinfo {author} {\bibfnamefont {N.}~\bibnamefont {Castagna}},
  \bibinfo {author} {\bibfnamefont {A.}~\bibnamefont {Hofer}}, \bibinfo
  {author} {\bibfnamefont {P.}~\bibnamefont {Knowles}}, \bibinfo {author}
  {\bibfnamefont {J.-L.}\ \bibnamefont {Schenker}}, \bibinfo {author}
  {\bibfnamefont {M.}~\bibnamefont {Kasprzak}}, \bibinfo {author}
  {\bibfnamefont {H.}~\bibnamefont {Saudan}}, \ and\ \bibinfo {author}
  {\bibfnamefont {A.}~\bibnamefont {Weis}},\ }\href {\doibase
  http://dx.doi.org/10.1063/1.3255041} {\bibfield  {journal} {\bibinfo
  {journal} {Applied Physics Letters}\ }\textbf {\bibinfo {volume} {95}},\
  \bibinfo {eid} {173701} (\bibinfo {year} {2009})}\BibitemShut {NoStop}%
\bibitem [{\citenamefont {Vrba}(1996)}]{Vrba}%
  \BibitemOpen
  \bibfield  {author} {\bibinfo {author} {\bibfnamefont {J.}~\bibnamefont
  {Vrba}},\ }in\ \href {\doibase 10.1007/978-94-011-5674-5_3} {{\selectlanguage
  {English}\emph {\bibinfo {booktitle} {SQUID Sensors: Fundamentals,
  Fabrication and Applications}}}},\ \bibinfo {series} {NATO ASI Series}, Vol.\
  \bibinfo {volume} {329},\ \bibinfo {editor} {edited by\ \bibinfo {editor}
  {\bibfnamefont {H.}~\bibnamefont {Weinstock}}}\ (\bibinfo  {publisher}
  {Springer Netherlands},\ \bibinfo {year} {1996})\ pp.\ \bibinfo {pages}
  {117--178}\BibitemShut {NoStop}%
\end{thebibliography}%

\appendix 

\setcounter{equation}{0}

\section{Gradiometer Figure of Merit}
\label{App:FOM}
\begin{figure}[h!]
	\centering
		\includegraphics[width = 0.9\linewidth]{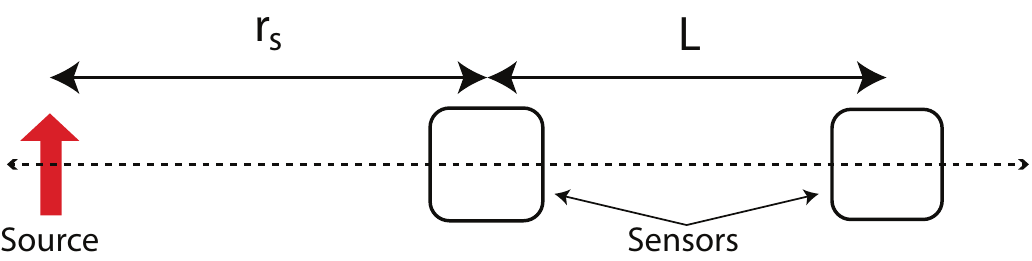}
	\caption{Configuration of source and magnetometers. The source, a fetal heart, is modeled to be a current dipole $I d\vec{l}$. In the cartoon, the dipole is oriented perpendicular to the axis of the magnetometers, but it should be noted the intra-heart currents that generate the detected fields flow along all three axes.}
	\label{fig:AppConfig}
\end{figure}

Consider a configuration as shown in Figure \ref{fig:AppConfig}. A magnetic field source is located a distance $r_s$ from a magnetometer, with a second magnetometer positioned at an additional distance of $L$ along the same axis. The source is assumed to generate a magnetic field with a power law $B(r_s) \propto (1/r)^p$.  For example, a point source magnetic dipole in a uniform medium would have a profile with $p = 3$. On the other hand, the field from a heart -- relevant for MCG -- is often modeled as a current dipole \cite{Cohen1979}, with  $p \sim 1.8-2.2$ .  

We express the output signal $S_i$ of each magnetometer as $S_i = M_i B(r_i)$ where $M$ is a response function and $B(r_i)$ is the field magnitude at a distance $r_i$ from the source. The difference signal $S_G$ is then given by:
\begin{align}
S_G = S_1-S_2 = M_1 B_1 - M_2 B_2.
\end{align}
The signal from a well-balanced (i.e. $M_1 \approx M_2 = M$) gradiometer in a magnetic field $B_i = B_0/r_i^p$ is
\begin{align}
\label{eq:gradsignal1A}
S_G \approx M(B_1 - B_2)	&= M \left( \frac{B_0}{r_s^p} - \frac{B_0}{\left(r_s+L\right)^p}\right) \nonumber \\
						&= \frac{-r_s^p + \left( r_s + L \right)^p}{\left( r_s+L \right)^p}MB_1 \nonumber \\
						&= \frac{g^p-1}{g^p}MB_1
\end{align}
where $g = 1 + L/r_s$, and $B_1$ is the magnetic field strength at the magnetometer closest to the source.

In practice, the magnetometers can have different amplitude and phase responses to identical magnetic field inputs. The resulting gradiometer consequently will acquire a residual sensitivity to correlated magnetic fields. The undesired residual sensitivity is inversely proportional to the common-mode rejection ratio $\xi$, where
\begin{align}
\label{eq:CMRR}
\xi 	&= \frac{\frac{1}{2}(S_{1corr} + S_{2corr})}{S_{1corr} - S_{2corr}} \nonumber\\
	&= \frac{\frac{1}{2}(M_1 + M_2)}{M_1 - M_2}.
\end{align}
$S_{icorr}$ is the measured signal from magnetometer $i$ due to an applied correlated field. In SQUID gradiometry \cite{Vrba}, $\xi$ is limited by the matching pickup coil areas or angular alignment of the two coils. In the case of atomic magnetometers, it is limited by a combination of the noise floor of the magnetometers and the calibration procedure.

In general, a measurement is made in a magnetically noisy environment with total noise $\delta B = \sqrt{(\delta B_u)^2 + (\delta B_c)^2}$.  $\delta B_u$ and $ \delta B_c$ are the uncorrelated and correlated sources of magnetic noise, respectively. The SNR for the two-channel gradiometer is then
\begin{align}
\label{eq:SNRG}
SNR_G 	&= \frac{\frac{g^p-1}{g^p}MB_1}{\sqrt{(M\delta B_{u1})^2 + (M\delta B_{u2})^2 + 2\left(M \delta B_{c}  \right / \xi)^2}} \nonumber \\
		&= \frac{g^p-1}{g^p}\frac{B_1}{\sqrt{2\left( \delta B_u \right)^2 + 2\left(\delta B_c / \xi  \right)^2}}.
\end{align}
where we have made the assumption that $\delta B_{u1} \approx \delta B_{u2}$ for the two magnetometers.

If, instead, the two magnetometers (again, with $M_1 \approx M_2 = M$) were not operated as a gradiometer, then the total signal is 
\begin{align}
\label{eq:magsignal1}
S_M 	&= S_1 + S_2 = M\left(B_1+B_2\right) \nonumber \\
		&= M \left( \frac{B_0}{r_s^p} + \frac{B_0}{\left(r_s+L\right)^p}\right) \nonumber \\
		&=\frac{g^p+1}{g^p}MB_1
\end{align}
Thus the signal-to-noise ratio for this two-channel magnetometer is
\begin{align} 
\label{eq:SNRM}
SNR_M &= \frac{g^p+1}{g^p}\frac{B_1}{\sqrt{2\left(\delta B_u\right)^2+2\left(\delta B_c \right)^2}}.
\end{align}

Comparing equations \ref{eq:SNRG} and \ref{eq:SNRM} illustrates the potential advantage of using a gradiometer, which we define as the gradiometer figure of merit $\mathcal{F}$.
\begin{align}
\mathcal{F} = \frac{SNR_G}{SNR_M}	&= \frac{g^p-1}{g^p+1}\frac{\sqrt{2\left(\delta B_u \right)^2 + 2\left(\delta B_c\right)^2}}{\sqrt{2\left(\delta B_u \right)^2 + 2\left( \delta B_c / \xi \right)^2}} \nonumber \\ 
                                                       	&= \frac{g^p-1}{g^p+1}\frac{\sqrt{\left(\frac{\delta B_u}{\delta B_c}\right)^2 + 1}}{\sqrt{\left(\frac{\delta B_u}{\delta B_c}\right)^2 + \xi^{-2}}}
\end{align}

Gradiometers are only useful if the total noise is dominated by correlated noise between the two channels. Under the assumption that $\delta B_u / \delta B_c \ll 1$,
\begin{align}
\label{eq:AFOM1}
\mathcal{F} \approx \frac{g^p-1}{g^p+1}\frac{1}{\sqrt{\left(\frac{\delta B_u}{\delta B_c}\right)^2 + \xi^{-2}}}.
\end{align}

This clearly demonstrates the the figure of merit depends on not only the noise ratio of $\delta B_u / \delta B_c$, but also on how that ratio compares to $\xi$. If $\left(\delta B_u/\delta B_c\right)^2 \ll \xi^{-2}$, immediate and potentially substantial benefits to the figure of merit (and therefore gradiometer performance) will be gained through an increase of $\xi$.  If, instead, $\left(\delta B_u/\delta B_c\right)^2 \gg \xi^{-2}$, further experimental efforts would be best focused elsewhere.

Taking $p = 2$ and $L = r_s$, for example, we obtain
\begin{equation}
\mathcal{F} \approx \frac{3}{5} \frac{1}{\sqrt{\left( \frac{\delta B_u}{\delta B_c} \right)^2 + \xi^{-2}}}.
\end{equation}

\section{SERF magnetometry}
\label{sec:AppSERFMag}
 For an ensemble of alkali atoms with total angular momentum F, electronic angular momentum $\mathbf{S}$ and electron spin polarization $\mathbf{P} = 2\mathbf{S}$, the SERF regime is where the alkali-alkali spin exchange rate is sufficiently high such that the atoms are described by a spin temperature distribution. We can describe the evolution of angular momentum with the Bloch equation:
\begin{equation}
   2\frac{\partial{\left< \mathbf{F} \right>}}{\partial t} = R\left( \mathbf{s}  -\left< \mathbf{P} \right> \right) + \mathbf{\Omega}\times \left< \mathbf{P} \right> - \Gamma \left< \mathbf{P} \right>
\label{eq:Bloch}
\end{equation}
R is the optical pumping rate, $\mathbf{s}$ the photon angular momentum vector. The magnetic field,  $\mathbf{B}= \mathbf{\Omega} / \gamma$ is the given in terms of the Larmor frequency and the gyro-magnetic ratio $\gamma$. $\Gamma$ is the total effective spin relaxation rate. In steady state, equation \ref{eq:Bloch} has solution:
\begin{equation}
    \left< \bf{P} \right>  = \frac{R}{\Gamma' (\Gamma'^2 + \Omega^2) }\left(\ \begin{array}{c}
         \Gamma'\Omega_y + \Omega_z\Omega_x \\
         -\Gamma'\Omega_x + \Omega_y\Omega_z \\
         -\Gamma'^2 + \Omega_z^2
    \end{array} \right) 
\end{equation}
where $\Gamma'=\Gamma+R$. In our design, we probe the x-component of $\bf{P}$ is probed via off-resonance Faraday rotation. 

In near zero field, we have $\mathbf{\Omega}\sim 0$. Consequently
\begin{equation}
    P_x \sim \frac{R}{\Gamma' + \Omega^2}\Omega_y 
\end{equation}

In the Z mode case, we apply a modulating field along the pumping direction, such that the total field in equation \ref{eq:Bloch} is now $$
{\bf{\Omega}} =\Omega_x \hat{x} + \Omega_y \hat{y} + (\Omega_z + \Omega_0 \cos(\omega_z t)) \hat{z}  
$$ yielding the results:
 \begin{equation}
     \begin{array}{ll}
          P_x(DC) & \approx  -\frac{P_z}{\Gamma'}\Omega_y J_0(u)^2 \\
          P_x(n\omega_z) & \approx  -\frac{P_z}{\Gamma'} 2J_0(u)J_n(u) 
          \begin{cases}
          -\Omega_x \sin(n\omega_z t), & n\text{ odd}\\
          -\Omega_y \cos(n\omega_z t), & n\text{ even}\\
          \end{cases}\\
     \end{array}
 \end{equation}
Consequently, we see that the $P_x$ signals demodulated at even(odd) harmonics of $\omega_z$ lead to sensitivity to $B_y$ ($B_z$). See \cite{WyllieThesis,DeLandThesis,Li} for more detailed description.
\end{document}